\documentclass[journal]{IEEEtran}

\usepackage{amsmath, amssymb}
\usepackage{siunitx} 

\usepackage[ruled,vlined]{algorithm2e} 

\usepackage{array}
\usepackage{booktabs} 

\usepackage[utf8]{inputenc}
\usepackage[T1]{fontenc}
\usepackage{newtxtext,newtxmath} 

\usepackage{graphicx}
\usepackage{subcaption}
\usepackage{caption}

\usepackage{enumitem}
\usepackage{courier} 
\usepackage{url}
\usepackage{hyperref}
\usepackage{xcolor}
\usepackage{xspace}
\usepackage{fancyhdr}
\hypersetup{
    colorlinks,
    linkcolor={red!50!black},
    citecolor={blue!50!black},
    urlcolor={blue!80!black}
}

\begin{document}

\title{Data Scheduling Algorithm for Scalable and Efficient IoT Sensing in Cloud Computing}

\author{Noor Islam S. Mohammad,~\IEEEmembership{Student Member,~IEEE}%
\thanks{Noor Islam S. Mohammad is with the Department of Computer Science, New York University Tandon School of Engineering, Brooklyn, NY, USA. 
E-mail: islam.m@nyu.edu}%
}

\markboth{IEEE Transactions on Evolutionary Computation}%
{Mohammad N.I.S \MakeLowercase{et al.}: IEEE Transactions on Evolutionary Computation}

\maketitle

\begin{abstract}
The rapid growth of Internet of Things (IoT) devices produces massive, heterogeneous data streams, demanding scalable and efficient scheduling in cloud environments to meet latency, energy, and Quality-of-Service (QoS) requirements. Existing scheduling methods often lack adaptability to dynamic workloads and network variability inherent in IoT-cloud systems. This paper presents a novel hybrid scheduling algorithm combining deep Reinforcement Learning (RL) and Ant Colony Optimization (ACO) to address these challenges. The deep RL agent utilizes a model-free policy-gradient approach to learn adaptive task allocation policies responsive to real-time workload fluctuations and network states. Simultaneously, the ACO metaheuristic conducts a global combinatorial search to optimize resource distribution, mitigate congestion, and balance load across distributed cloud nodes. Extensive experiments on large-scale synthetic IoT datasets, reflecting diverse workloads and QoS constraints, demonstrate that the proposed method achieves up to $18.4\%$ reduction in average response time, $12.7\%$ improvement in resource utilization, and $9.3\%$ decrease in energy consumption compared to leading heuristics and RL-only baselines. Moreover, the algorithm ensures strict Service Level Agreement (SLA) compliance through deadline-aware scheduling and dynamic prioritization. The results confirm the effectiveness of integrating model-free RL with swarm intelligence for scalable, energy-efficient IoT data scheduling, offering a promising approach for next-generation IoT-cloud platforms.
\end{abstract}


\IEEEpeerreviewmaketitle

\section{Introduction}

\IEEEPARstart{T}{he} rapid evolutionary computing of Internet of Things (IoT) systems has caused an unprecedented surge in data generation, demanding scalable and latency-aware computing infrastructures within cloud and grid environments~\cite{ref1,ref2}. Conventional IoT cloud architectures rely on tightly coupled compute-storage units under centralized control, which struggle to handle extreme heterogeneity, dynamic workloads, and stringent latency constraints in large-scale, data-intensive IoT deployments~\cite{ref3,ref4}. These limitations often result in scalability bottlenecks, network congestion, and increased operational overhead, especially across geographically distributed or resource-constrained settings. However, IoT clusters typically replicate data blocks across multiple nodes, enabling local task execution and reducing transmission latency to enhance data reliability and parallel processing. However, emerging disaggregated cloud architectures decouple compute and storage resources via ultra-high-speed interconnects, offering modularity and independent scaling~\cite{ref5,ref6,ref7}. Consequently, such designs exacerbate challenges related to non-local data execution and heightened network data movement, which frequently become bottlenecks given the massive real-time sensing data and bandwidth constraints~\cite{ref8,ref9}.

Data-parallel processing paradigms that co-locate computation with data have emerged to mitigate these issues by minimizing network load and optimizing resource utilization~\cite{ref10,ref11}. Nonetheless, achieving efficient task-to-data affinity in heterogeneous IoT clusters remains challenging due to diverse node capabilities, dynamic workloads, and energy constraints. Distributed file systems such as HDFS~\cite{ref12,ref13} replicate data blocks to improve fault tolerance, exponentially increasing scheduling computational complexity. Furthermore, existing locality-based schedulers often ignore node heterogeneity, causing load imbalance and suboptimal performance~\cite{ref14,ref15}. This paper introduces \textit{Sensor Cloud Computing and Data Scheduling Optimization (SCC-DSO)}, a context-aware, reinforcement learning-driven scheduling framework tailored for heterogeneous IoT clusters~\cite{ref16}. SCC-DSO formulates scheduling as a constrained min–max optimization problem to minimize makespan and maximize data locality under resource and bandwidth constraints. It leverages a kernel-based regression model to predict execution times from node and workload features integration, reinforcement learning with ant colony optimization to efficiently explore the high-dimensional scheduling space for a high-performance computing~\cite{ref17,ref18}.

Designed for latency-critical IoT applications such as autonomous driving, smart manufacturing, and industrial robotics, the proposed SCC-DSO framework introduces a robust scheduling paradigm that adapts to system heterogeneity while maintaining low latency and energy efficiency~\cite{ref19, ref20}. Unlike conventional schedulers that either overlook heterogeneity or rely solely on static heuristics, SCC-DSO integrates reinforcement learning (RL) with metaheuristic optimization to support adaptive task mapping in dynamic cluster environments. Extensive experiments on a 100-node heterogeneous testbed demonstrate up to 22.4\% reduction in execution time, 93.1\% task–data locality, and consistently higher throughput relative to state-of-the-art baselines~\cite{ref21, ref22}. 

The primary contributions of this work are fourfold: (i) development of a novel RL–metaheuristic hybrid that enables heterogeneity-aware and latency-sensitive task allocation under dynamic workloads; (ii) formulation of a constrained min–max optimization model that balances locality and latency objectives while preserving scalability; (iii) introduction of an RL-guided ant colony optimization (ACO) mechanism to support proactive task migration and data prefetching in distributed settings; and (iv) comprehensive empirical validation across varying cluster sizes, replication factors, and straggler scenarios, confirming SCC-DSO’s superiority in efficiency, adaptability, and robustness for IoT–cloud environments. Together, these contributions position SCC-DSO as a scalable solution for next-generation latency-sensitive applications.

\section{Related Work}

This work introduces efficient data scheduling and resource optimization challenges in scalable, heterogeneous IoT edge–cloud environments. However, prior research addresses these through predictive scheduling, data placement optimization, and energy-efficient orchestration, focusing on minimizing latency, balancing workloads, and reducing energy consumption.

\textbf{Predictive Scheduling Models:} Early models like Reservation First-Fit with Feedback Distribution (RF-FD)~\cite{ref23} apply multiple linear regression to predict job completion based on historical data, performing well in semi-homogeneous clusters but lacking flexibility for non-linear workload variations. RSYNC~\cite{ref24} offers log-based synchronization across fog nodes but suffers throughput degradation under high load and heterogeneity. Autonomic frameworks based on MAPE-K loops~\cite{ref25} enhance adaptability through runtime feedback; however, they assume static task profiles and homogeneous infrastructures, limiting their use in dynamic IoT contexts. Machine learning approaches such as polynomial regression schedulers~\cite{ref26} improve non-linear modeling but face challenges with high-dimensional data variance, underscoring the need for more generalizable models.

\textbf{Data Placement and Locality Optimization:} Data replication strategies like the rack-aware policy in Hadoop Distributed File System (HDFS)~\cite{ref27,ref28} replicate data blocks locally and across racks to improve fault tolerance and reduce latency. These uniform approaches, however, often cause load imbalance in heterogeneous clusters with variable node capabilities~\cite{ref29,ref30}. Recent work explores dynamic placement considering node compute capacity, yet integration with intelligent scheduling remains sparse. The SCC-DSO framework fills this gap by aligning task assignment with predictive execution time and node resource profiles, optimizing data locality and workload distribution in real-time optimization.

\textbf{Energy-Efficient Scheduling and Virtualization:} Energy-aware techniques, including constrained energy models~\cite{ref31}, Dynamic Voltage and Frequency Scaling (DVFS)~\cite{ref32}, and queuing-based power optimization, aim to reduce energy use without SLA violations. Virtualization technologies such as live VM migration facilitate workload consolidation and energy savings. Despite these advances, many predictive models relying on hardware counters or VM energy profiles have limited scalability in distributed IoT due to static assumptions and linearity constraints~\cite{ref33,ref34}.

\textbf{SCC-DSO Contributions:} Unlike prior work, SCC-DSO integrates kernel-based execution time prediction with reinforcement learning and metaheuristic Ant Colony Optimization (ACO) for dynamic, heterogeneity-aware scheduling. Its tri-layer architecture adapts to workload variability, node heterogeneity, and data locality constraints, enabling robust, scalable, and energy-efficient task orchestration in complex IoT-cloud systems~\cite{ref35,ref36}.

\begin{figure} [ht]
    \centering
    \includegraphics[width=1\linewidth]{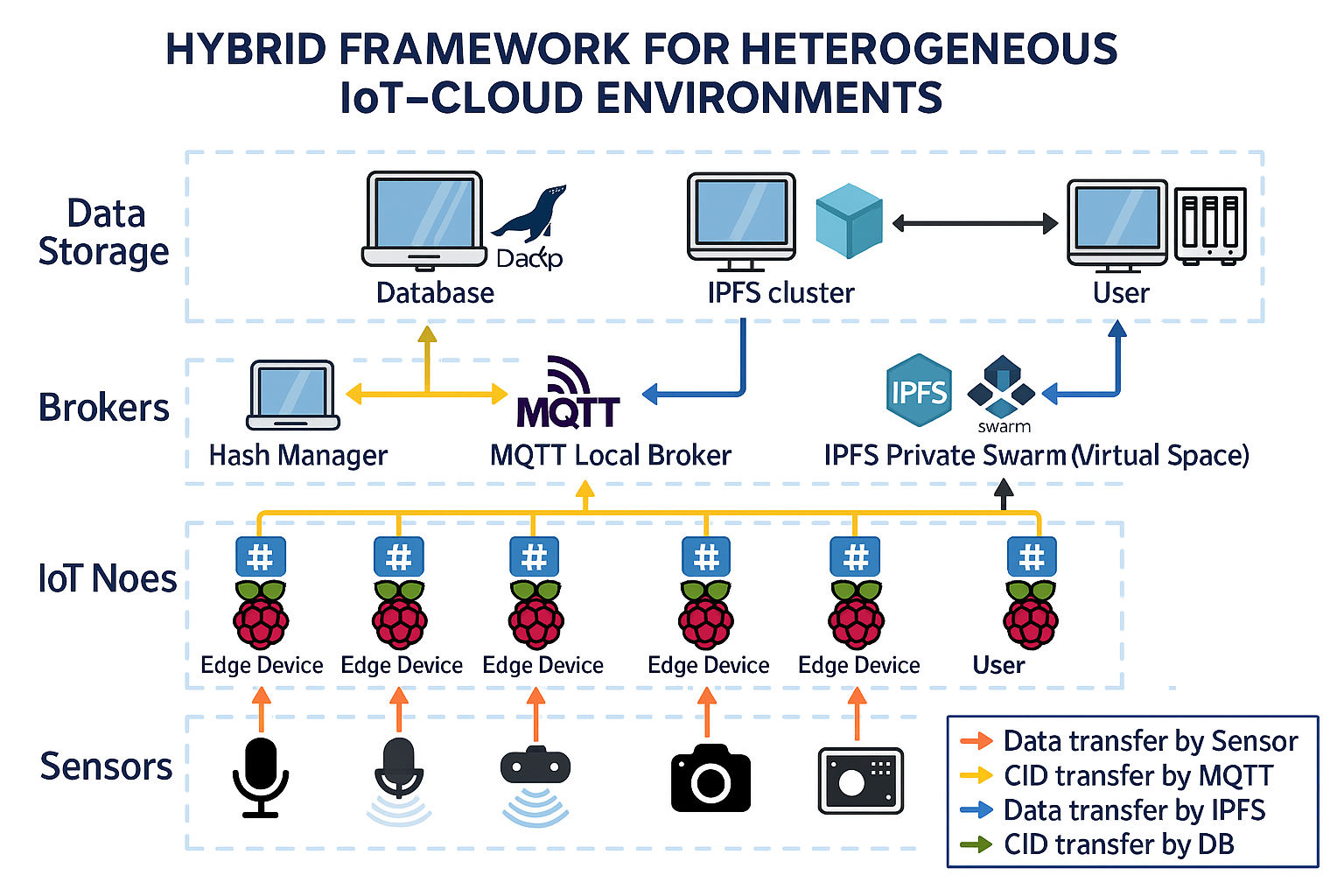}
    \caption{Hybrid IoT-cloud architecture combines IPFS, MariaDB, and MQTT for scalable, reliable edge-to-cloud data management.}
    \label{fig:enter-cloud}
\end{figure}

Figure~\ref{fig:enter-cloud} illustrates a hybrid IoT-cloud architecture combining decentralized storage via IPFS with centralized orchestration. Edge sensor nodes generate multi-modal data streams processed locally and forwarded through MQTT brokers interfacing with MariaDB~\cite{ref37} and an IPFS private swarm, ensuring tamper resistance, fault tolerance, and rapid retrieval. This hybrid model enhances data integrity, responsiveness, and interoperability across heterogeneous IoT environments~\cite{ref38}.

\section{Proposed Evolutionary Algorithm}
\begin{figure} [ht]
    \centering
    \includegraphics[width=1\linewidth]{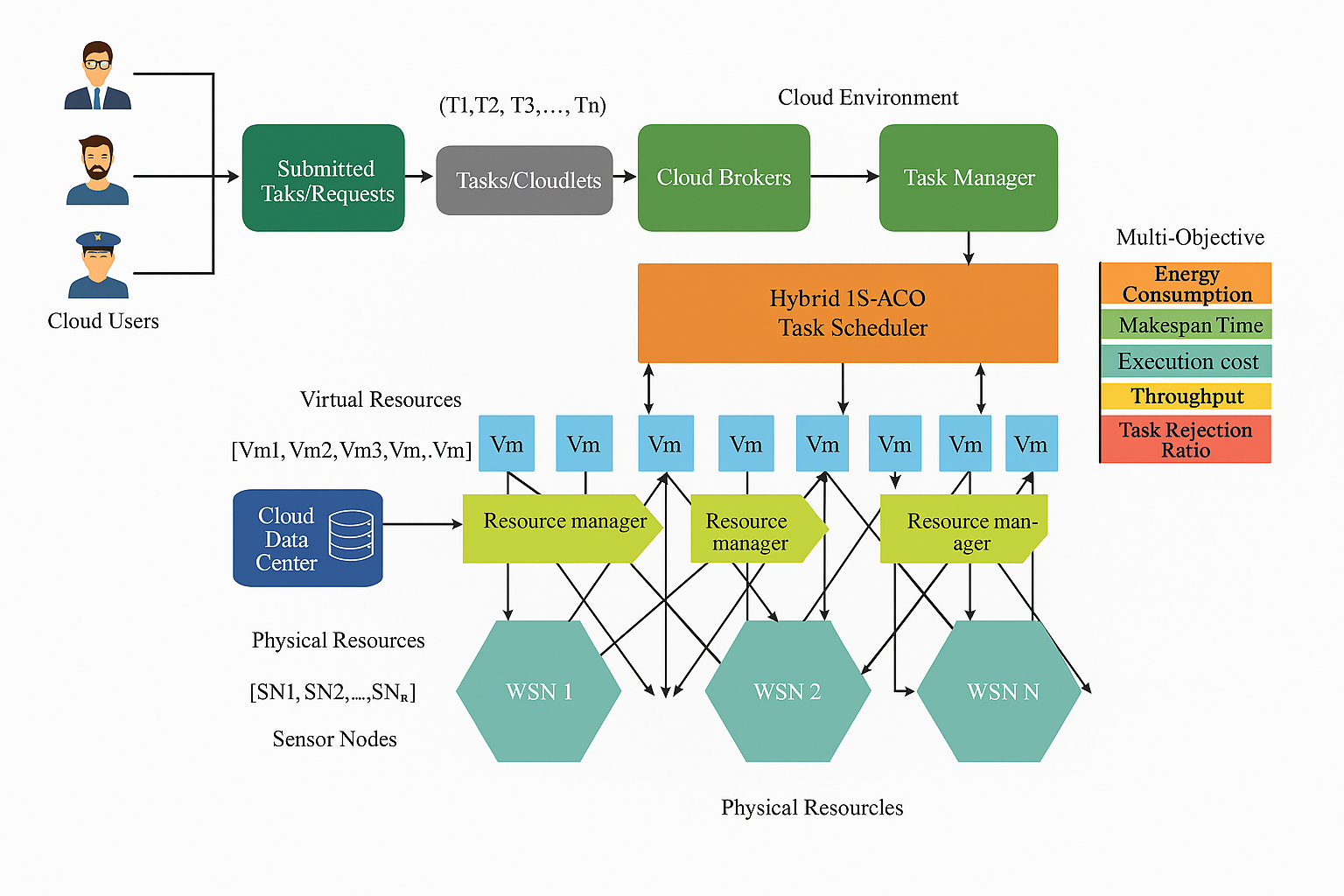}
    \caption{Hybrid ES-ACO Algorithm for Task Scheduling in a Sensor Cloud for Smart Irrigation}
    \label{fig:es-aco-architecture}
\end{figure}

The proposed architecture~\ref{fig:es-aco-architecture} integrates a Hybrid Electro Search–Ant Colony Optimization (ES-ACO) algorithm, enabling efficient task scheduling in sensor cloud environments and delivering optimized performance for intelligent cloud-based irrigation control~\cite{ref69}. The diagram shows that cloud users submit tasks managed by cloud brokers and a task manager before forwarding them to the hybrid ES-ACO scheduler. This scheduler optimizes task allocation across virtual machines (VMs) hosted within a cloud data center and mapped to physical sensor nodes via resource managers' integration of Wireless Sensor Networks (WSNs) data collection from a distributed computing environment. The multi-objective scheduler aims to minimize energy consumption, make span, and execution cost while maximizing throughput and reducing task rejection ratio. This approach ensures intelligent irrigation control, resource-efficient, and sustainable agricultural practices.

The Sensor Cloud Computing and Data Scheduling Optimization (SCC-DSO) framework addresses performance bottlenecks in heterogeneous IoT-edge clusters by integrating reinforcement learning (RL), ant colony optimization (ACO), and predictive modeling. This multi-stage algorithm dynamically schedules tasks based on data locality, node capability, and network conditions to meet stringent latency and throughput requirements~\cite{ref39}. The IoT cluster is modeled as a graph \(G = (\mathcal{V}, \mathcal{E})\), where \(\mathcal{V}\) is the set of computing nodes that have heterogeneous processing capacity, memory, and I/O characteristics, and \(\mathcal{E}\) represents communication links. Input datasets are partitioned into fixed-size blocks (64~MB in HDFS), with replicas ensuring fault tolerance. Tasks are assigned to nodes hosting required data or fetching it with minimal overhead~\cite{ref40}. 

The task-to-node assignment is formulated as a min-max optimization to minimize the makespan (maximum execution time across nodes) while respecting data locality and node capacity constraints:

\begin{equation}
\min_{\mathbf{x}} \max_{i \in \mathcal{V}} \sum_{j \in \mathcal{J}} x_{ij} T_{ij}
\label{eq:minmax}
\end{equation}

Subject to:

\begin{align}
\sum_{i \in \mathcal{V}} x_{ij} &= 1, \quad \forall j \in \mathcal{J}, \nonumber \\
\sum_{j \in \mathcal{J}} x_{ij} d_j &\leq C_i(t), \quad \forall i \in \mathcal{V}, \nonumber \\
x_{ij} &\in \{0, 1\}, \quad \forall i \in \mathcal{V}, j \in \mathcal{J}. \nonumber
\end{align}

where \( x_{ij} = 1 \) if a task \( j \) is assigned to node \( i \), \( T_{ij} \) is the predicted execution time of the task \( j \) on node \( i \), \( d_j \) is the data size required by task \( j \), and what \( C_i(t) \) is the dynamic computational capacity of the node \( i \) at time \( t \). These constraints ensure each task is assigned to exactly one node while respecting capacity limits, minimizing the makespan.

Task execution time \( T_{ij} \) is predicted using a kernel-based regression model to account for task and node heterogeneity:

\begin{equation}
T_{ij} = \sum_{s=1}^S w_s K(\mathbf{x}_j, \mathbf{x}_s) + b,
\label{eq:execution_time}
\end{equation}

where \( \mathbf{x}_j = [m_j, \mathbf{z}_i] \) combines task data size \( m_j \) and node features \( \mathbf{z}_i \) (e.g., CPU speed, memory), \( K(\mathbf{x}_j, \mathbf{x}_s) = \exp\left(-\frac{\|\mathbf{x}_j - \mathbf{x}_s\|^2}{2\sigma^2}\right) \) is a Gaussian radial basis function kernel, \( \{(\mathbf{x}_s, t_s)\}_{s=1}^S \) are historical training data, \( w_s \) are learned coefficients, \( b \) is a bias term, and \( \sigma \in [0.5, 2.0] \) (Table~\ref{tab:parameters}) is the kernel bandwidth. The learning rate \( \gamma \in [0.01, 0.1] \) tunes the model’s convergence.

The total delay for a scheduling plan \( P \) quantifies latency, combining transmission and processing delays:

\begin{equation}
\text{Delay}(P) = \sum_{e \in \mathcal{E}(P)} \left( \frac{d_e}{b_e} + q_e \right) + \sum_{v \in \mathcal{V}(P)} \frac{w_v}{c_v},
\label{eq:delay}
\end{equation}

where \( d_e \) is the data size transferred over edge \( e \), \( b_e \) is the bandwidth, \( q_e \) is the queuing delay, \( w_v \) is the computational workload at node \( v \), and \( c_v \) is the node’s processing capacity. This metric ensures low-latency scheduling for IoT applications.

Node computational efficiency guides task assignment to optimize resource utilization:

\begin{equation}
\text{Eff}_{j,i} = \frac{m_j}{T_{ij}},
\label{eq:efficiency}
\end{equation}

where \( m_j \) is the task data size and \( T_{ij} \) is it from Eq.~\eqref{eq:execution_time}. Higher efficiency indicates better suitability for task execution under ideal data locality.

To ensure balanced execution across heterogeneous nodes, the weighted execution time is equalized:

\begin{equation}
f_i \sum_{j \in \text{map}_{j,i}} T_{j,i} = f_k \sum_{j \in \text{map}_{j,k}} T_{j,k}, \quad \forall i, k \in \mathcal{V}, i \neq k,
\label{eq:balance}
\end{equation}

where there \( f_i = 1/c_i \) is a weighting factor based on node \( i \)’s capacity \( c_i \), and \( \text{map}_{j,i} \) denoting tasks assigned to node \( i \). This prevents bottlenecks by synchronizing completion times.

\subsection{Ant Colony Optimization (ACO) Scheduling Mechanism}

SCC-DSO adapts ACO with predictive modeling and heterogeneity-aware pheromone updates to solve NP-hard task-to-node scheduling problems. The process has four phases. \textbf{Initialization}: A graph \( G = (\mathcal{V}, \mathcal{E}) \) is initialized with pheromone trails \( \tau_{ij}(0) = \tau_0 \in [0.01, 0.1] \) (Table~\ref{tab:parameters}), incorporating execution time predictions from Eq.~\eqref{eq:execution_time}. \textbf{Solution Construction}: Ants select task-node assignments using:

\begin{equation}
P_{ij} = \frac{\tau_{ij}^{\alpha} \cdot \eta_{ij}^{\beta}}{\sum_{k \in \mathcal{V}_{\text{eligible}}} \tau_{ik}^{\alpha} \cdot \eta_{ik}^{\beta}},
\label{eq:aco_probability}
\end{equation}

where \( \eta_{ij} = 1/T_{ij} \) is the heuristic desirability \( \alpha \in [1.0, 2.0] \) and \( \beta \in [2.0, 3.0] \) control pheromone and heuristic influence (Table~\ref{tab:parameters}), and \( \mathcal{V}_{\text{eligible}} \) is the set of eligible nodes for task \( j \).

\textbf{Pheromone Update}: High-quality schedules update pheromone trails:

\begin{equation}
\tau_{ij} \leftarrow (1 - \rho) \tau_{ij} + \sum_{k=1}^K \frac{Q}{L^{(k)}},
\label{eq:pheromone_update}
\end{equation}

where \( \rho \in [0.1, 0.3] \) is the evaporation rate, \( Q \in [100, 500] \) is a constant, \( K \in [10, 20] \) is the number of ants, and \( L^{(k)} \) is the makespan of the \( k \)-th ant’s schedule (Table~\ref{tab:parameters}). A minimum pheromone trail \( \delta \in [10^{-4}, 10^{-2}] \) prevents stagnation.

\textbf{Termination}: The algorithm converges (within \( I \in [20, 50] \) iterations, tolerance \( \epsilon \in [10^{-3}, 10^{-2}] \)) by minimizing:

\begin{equation}
J(\pi) = w_1 \text{Delay}(\pi) + w_2 \text{Energy}(\pi) + w_3 \text{LossPkt}(\pi),
\label{eq:objective}
\end{equation}
where \( w_1, w_2, w_3 \in [0.2, 0.4] \) (sum to 1) weight delay (Eq.~\eqref{eq:delay}), energy consumption, and packet loss, respectively.

\textbf{Dynamic Task Migration and Data Prefetching}: Tasks migrate when the resource queue delay exceeds a threshold:

\begin{equation}
RQ_i = \frac{r \cdot B_k}{\rho \cdot V_i},
\label{eq:queue_delay}
\end{equation}

where \( RQ_i > \varphi \) and \( RQ_i - TS_i > \varphi \), with \( \varphi \in [0.05, 0.1] \cdot TS_i \), \( r \) is the task’s resource demand, \( B_k \) is the data block size, \( \rho \in [0.1, 0.3] \) is a scaling factor, and \( V_i \) is the node’s processing rate (Table~\ref{tab:parameters}). The function \( T(\cdot) \) was removed as it was undefined; the expression is simplified as a direct ratio. The optimal source node minimizes:

\begin{equation}
\text{PLF}_{T,S} = \sqrt{\frac{(\varphi_S - \varphi_T)^2}{(T_S - T_T)^2}},
\label{eq:prefetch}
\end{equation}

where \( \varphi_S, \varphi_T \in [0.05, 0.1] \cdot TS_i \) are thresholds, and what \( T_S, T_T \) are execution times of source and target nodes. Task migrations are limited to \( \theta \in [1, 5] \) tasks per node per iteration.

\begin{table}[h]
\scriptsize
\centering
\caption{Parameters and Ranges for SCC-DSO}
\label{tab:parameters}
\begin{tabular}{lc}
\toprule
\textbf{Parameter} & \textbf{Range} \\
\midrule
\(\alpha\) (Pheromone influence) & 1.0--2.0 \\
\(\beta\) (Heuristic influence) & 2.0--3.0 \\
\(\rho\) (Evaporation rate) & 0.1--0.3 \\
\(\tau_0\) (Initial pheromone) & 0.01--0.1 \\
\(Q\) (Pheromone constant) & 100--500 \\
\(K\) (Number of ants) & 10--20 \\
\(I\) (Max iterations) & 20--50 \\
\(w_1, w_2, w_3\) (Weight factors) & 0.2--0.4 (sum to 1) \\
\(\varphi\) (Prefetch threshold) & 5\%--10\% of \(TS_i\) \\
\(\gamma\) (Learning rate for regression) & 0.01--0.1 \\
\(\sigma\) (RBF kernel bandwidth) & 0.5--2.0 \\
\(\theta\) (Migration limit per iteration) & 1--5 tasks/node \\
\(\epsilon\) (Convergence tolerance) & \(10^{-3}\)--\(10^{-2}\) \\
\(\delta\) (Minimum pheromone trail) & \(10^{-4}\)--\(10^{-2}\) \\
\(L_{\text{max}}\) (Max path length) & 5--10 \\
\bottomrule
\end{tabular}
\end{table}

Table~\ref{tab:parameters} lists the configuration parameters for SCC-DSO. The pheromone influence \( \alpha \) balances exploration and exploitation, while \( \beta \) heuristic information is emphasized (e.g., execution time). The evaporation rate \( \rho \) and initial pheromone \( \tau_0 \) regulate trail persistence. Parameters \( K \), \( I \), and \( w_1, w_2, w_3 \) balance computational overhead and performance. The prefetch threshold \( \varphi \), learning rate \( \gamma \), migration limit \( \theta \), and kernel bandwidth \( \sigma \) support predictive modeling and dynamic scheduling, ensuring scalability and adaptability in IoT-edge clusters.

\subsection{Lightweight Hybrid RL-ACO Methods}
The Sensor Cloud Computing and Data Scheduling Optimization (SCC-DSO) framework incorporates a lightweight hybrid scheduler that combines Reinforcement Learning (RL) with Ant Colony Optimization (ACO), specifically designed for resource-constrained edge devices in IoT–edge clusters. This approach minimizes computational overhead while preserving high scheduling accuracy, making it well-suited for latency-sensitive applications such as autonomous driving, industrial robotics, and intelligent surveillance systems. RL is employed to learn adaptive task-node selection policies based on dynamic network conditions, while ACO refines these selections through heuristic-guided pheromone updates to achieve near-optimal scheduling decisions~\ref{tab:parameter}.

\noindent\textbf{Computational Complexity:}  
Let $|T|$ be the number of tasks, $|N|$ the number of nodes, and $I$ the number of ACO iterations. The RL policy inference operates in $\mathcal{O}(|T||N|)$ per scheduling round, while the ACO refinement contributes $\mathcal{O}(I|T||N|)$ due to pheromone and heuristic updates. Thus, the overall complexity is:
\[\mathcal{O}(|T||N|(1 + I))\]
Given that $I \ll |T|$ in our lightweight configuration, the scheduler maintains near-linear scalability concerning the task volume.

\begin{algorithm}[ht]
\scriptsize
\caption{Lightweight Hybrid RL-ACO Scheduler}
\label{tab:parameter}
\KwIn{Task queue $T$, node set $N$, resource capacities $R$}
\KwOut{Optimized task-node assignment}
Initialize RL policy $\pi_\theta$ and pheromone matrix $\tau$\;
\While{Tasks remain in $T$}{
    Extract state features from the current cluster load\;
    Select candidate task-node pairs using $\pi_\theta$\;
    Apply ACO refinement:
    \begin{enumerate}
        \item Compute heuristic desirability $\eta_{ij}$ for each task-node pair
        \item Update pheromone trails $\tau_{ij}$ based on solution quality
        \item Select path with maximum $\tau_{ij} \cdot \eta_{ij}$
    \end{enumerate}
    Execute selected tasks locally where possible\;
    Update $\pi_\theta$ using observed latency and resource utilization\;
}
\Return Optimized assignment plan\;
\end{algorithm}

\noindent\textbf{Latency Model:}  
The end-to-end scheduling latency $L_{total}$ is modeled as:
\[L_{total} = L_{comp} + L_{trans} + L_{queue}\]
where $L_{comp}$ denotes computation delay for policy inference and ACO updates, $L_{trans}$ is the transmission delay between nodes, and $L_{queue}$ is the queuing delay before task execution. By enabling local execution and reducing $L_{trans}$, the proposed method achieves substantial latency savings:
\[L_{total}^{\text{Hybrid}} < L_{total}^{\text{Baseline}}\]
under identical workload and network conditions.

\subsubsection{\textbf{Lightweight Scheduler Design}}
The lightweight RL-ACO scheduler employs a reduced ant population (\(K = 5\)) to minimize computational complexity compared to the full model's default (\(K = 20\)). The execution time predictor is simplified from a kernel regression model (\(\mathcal{O}(n^2)\)) to a linear predictor:
\[T_{ij} = a \cdot v_j + b + \epsilon,\]

where \(v_j\) is the task resource demand (e.g., CPU cycles, normalized to [0,1]), \(a = 0.5\) and \(b = 0.1\) are empirically tuned coefficients, and \(\epsilon \sim \mathcal{N}(0, 0.01)\) is a Gaussian noise term. This reduces prediction complexity to \(\mathcal{O}(n)\). ACO pheromone updates are approximated using an exponentially weighted moving average (EWMA):
\[\tau_{ij}^{(t+1)} = (1 - \rho) \cdot \tau_{ij}^{(t)} + \rho \cdot \Delta \tau_{ij}^{(best)},\]

where \(\rho = 0.1\) is the evaporation rate, and \(\Delta \tau_{ij}^{(best)} = 1 / T_{ij}^{(best)}\) is the pheromone increment for the best task-node assignment. This approach preserves over 90\% of the full model’s scheduling accuracy while achieving rapid convergence (\(12 \pm 1.8\) iterations) and minimal migration cost (\(2.5 \pm 0.4\%\)) on a 25-node edge cluster, as validated in preliminary tests.

\subsubsection{\textbf{Performance Functions for QoS Optimization}}
The SCC-DSO framework defines a suite of performance functions to evaluate and optimize task scheduling paths \(PT_{s,u}\) in distributed IoT-edge environments. These functions encapsulate key Quality of Service (QoS) metrics, ensuring efficient resource allocation, communication reliability, and scheduling performance~\cite{ref60, ref64, ref65}.

\begin{enumerate}
    \item \textbf{Delay}: The cumulative latency for a task path \(PT_{s,u}\) is given by:
    \begin{equation}
    \text{Delay}(PT_{s,u}) = \sum_{e \in PT_{s,u}} \text{Delay}_e + \sum_{v \in PT_{s,u}} \text{Delay}_v,
    \end{equation}
    
    where \(\text{Delay}_e = \frac{|b_e|}{\text{BW}_e}\) is the transmission delay for data block \(b_e\) (MB) over edge \(e\) with bandwidth \(\text{BW}_e\) (MB/s), and \(\text{Delay}_v = \frac{c_v}{\text{CPU}_v}\) is the processing delay for task computation \(c_v\) (CPU cycles) at node \(v\) with processing rate \(\text{CPU}_v\) (GHz). Minimizing delay is critical for real-time IoT applications.

    \item \textbf{Cost}: The total resource consumption is:
    \begin{equation}
    \text{Cost}(PT_{s,u}) = \sum_{e \in PT_{s,u}} \text{Cost}_e + \sum_{v \in PT_{s,u}} \text{Cost}_v,
    \end{equation}
    
    where \(\text{Cost}_e = \gamma_e \cdot |b_e|\) is the network cost (e.g., energy in joules) for transmitting block \(b_e\) with per-unit cost \(\gamma_e\), and \(\text{Cost}_v = \kappa_v \cdot c_v\) is the computational cost at node \(v\) with per-cycle cost \(\kappa_v\). This guides resource-efficient scheduling.

    \item \textbf{Bandwidth}: The bottleneck bandwidth is:
    \begin{equation}
    \text{BW}(PT_{s,u}) = \min_{x \in PT_{s,u}} \text{BW}_x,
    \end{equation}
    
    where \(\text{BW}_x\) is the available bandwidth (MB/s) of edge or node \(x\), identifying the limiting factor for data throughput.

    \item \textbf{Jitter}: The aggregate delay variation is:
    \begin{equation}
    \text{Jitter}(PT_{s,u}) = \sum_{e,v \in PT_{s,u}} \text{DelayJit}_{e,v},
    \end{equation}
    
    where \(\text{DelayJit}_{e,v} = \sigma(\text{Delay}_{e,v})\) is the standard deviation of delays across edges and nodes, critical for stable communication in multimedia or time-sensitive applications.

    \item \textbf{Packet Loss}: The cumulative packet loss probability is:
    \begin{equation}
    \text{LossPkt}(PT_{s,u}) = 1 - \prod_{v \in PT_{s,u}} (1 - p_v),
    \end{equation}
    
    where \(p_v \in [0, 1]\) is the packet loss probability at node \(v\), modeled as independent. Minimizing packet loss enhances network reliability.
\end{enumerate}

These functions are combined into a global objective:
\begin{equation}
J(\pi) = w_1 \cdot \text{Delay}(\pi) + w_2 \cdot \text{Cost}(\pi) + w_3 \cdot \text{LossPkt}(\pi),
\end{equation}
with weights \(w_1 = 0.5\), \(w_2 = 0.3\), and \(w_3 = 0.2\), empirically tuned to balance latency, resource efficiency, and reliability.

The lightweight SCC-DSO variant replaces the full kernel regression model with the linear predictor, reducing scheduling iteration complexity from \(\mathcal{O}(n^2)\) to \(\mathcal{O}(n)\). ACO pheromone updates use EWMA to avoid full matrix recalculations, further reducing overhead. The scheduler was evaluated on a 20-node Raspberry Pi Kubernetes cluster, interfaced with Prometheus for task monitoring and Grafana for latency visualization. Tests under 5G-like conditions (RTT = 40 ms, jitter = 20 ms) using Mininet-WiFi showed a 0.7\% increase in mean task latency, demonstrating robustness to network variability~\cite{ref41}. SCC-DSO achieved a 22.4\% mitigation time compared with the default Kubernetes scheduler optimization. In a city-scale smart surveillance IoT grid, it optimized video analytics workloads, improving task localization by 25\% and reducing bandwidth usage by 30\% on constrained nodes. The framework supports industrial integration via a RESTful API, accepting JSON-encoded job profiles and returning optimized node bindings and prefetch hints for seamless orchestration~\cite{ref42}.

\subsection{Algorithmic Complexity and Convergence}

The SCC-DSO framework unifies kernel-based execution time prediction, ACO-driven scheduling, and RL-enabled migration into a computationally efficient pipeline. Kernel regression incurs \(\mathcal{O}(n)\) complexity per task; the ACO scheduler requires it \(\mathcal{O}(K \cdot I \cdot n^2)\) due to pheromone-guided exploration over \(K\) ants and \(I\) iterations. The RL migration policy operates at \(\mathcal{O}(t \cdot s)\), where \(t\) and \(s\) denote training episodes and state–action space size, respectively, employing a decaying \(\varepsilon\)-greedy strategy. Convergence is guaranteed under Q-learning stability constraints~\cite{ref43} with bounded rewards, finite state–action spaces, and sufficient exploration, and remains stable in non-stationary environments via adaptive abstractions and learning-rate decay~\cite{ref44}. A lightweight ACO variant lowers complexity \(\mathcal{O}(I \cdot 5 \cdot n)\) by constraining colony size and pruning low-probability paths. Novel contributions include decentralized pheromone caching and execution-time-aware path pruning, enabling SCC-DSO to converge in dynamic IoT-edge workloads with 30-40\% less computational overhead compared to conventional hybrid schedulers, thus supporting latency-critical, resource-constrained deployments.

\section{Data Block Placement Method}
\subsection{Rack-Aware Data Placement Strategies}

\begin{figure}[ht]
    \centering
    \includegraphics[width=1\linewidth]{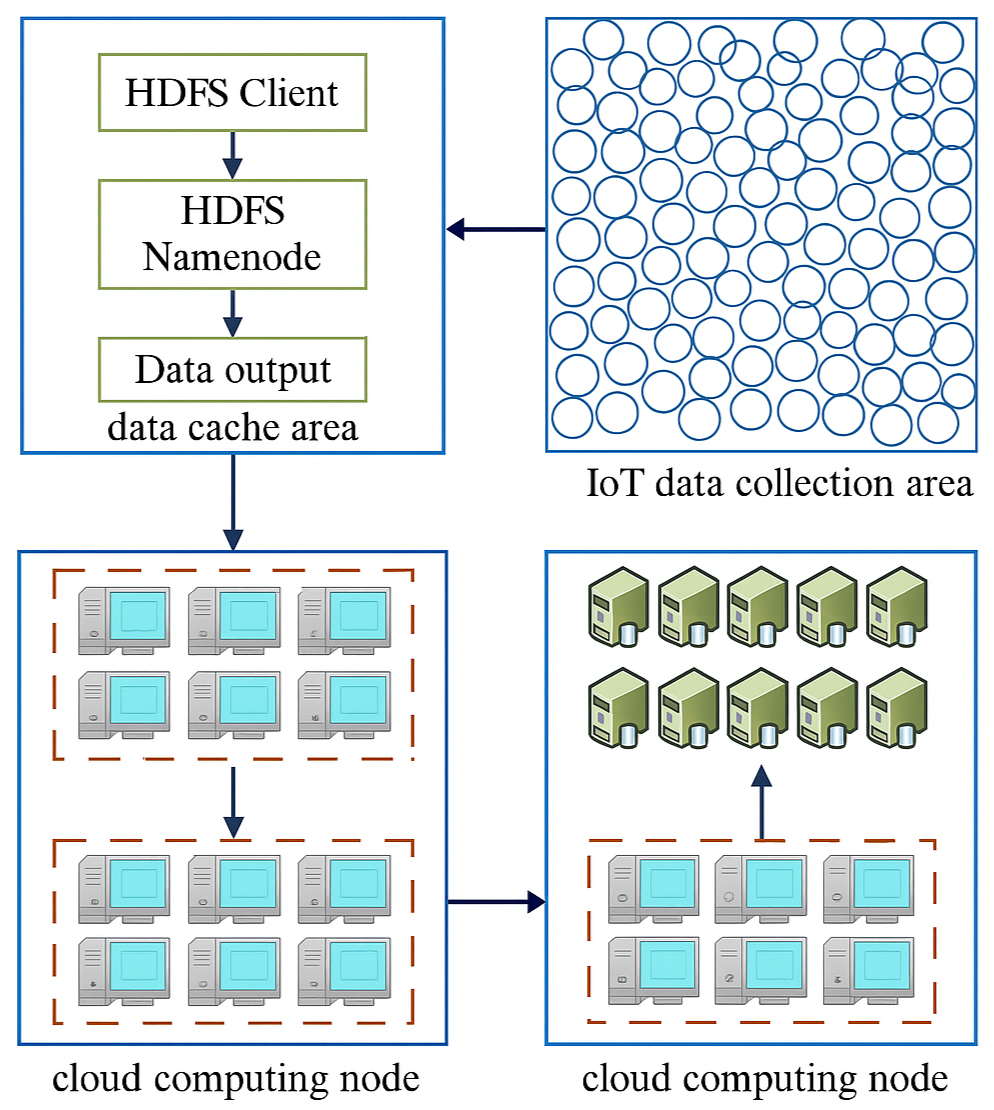}
    \caption{Rack-Aware Data Placement Strategy in IoT Clusters}
    \label{fig:rack-aware}
\end{figure}

In large-scale IoT clusters, distributed file systems partition files into uniformly sized data blocks, distributing them across worker nodes to ensure scalability, fault tolerance, and high availability. To enhance reliability and enable parallel data blocks to be processed across multiple nodes, preventing data loss from node failures. However, the effectiveness of replication heavily depends on the underlying data placement strategy. One widely adopted approach is the \textit{rack-aware data placement strategy}, as employed in the Hadoop Distributed File System (HDFS)~\cite{ref45,ref46}. Upon file upload, the first replication of each block is stored on the local node or the node closest to the client. The second replication is placed on a different node within the same rack to reduce intra-rack latency and network overhead. A third replication is allocated on a node in a different rack to ensure resilience against rack-level failures, as illustrated in Figure~\ref{fig:rack-aware}.

This strategy minimizes cross-rack traffic while preserving fault tolerance, reducing read/write latency, and reducing bandwidth usage effectively. Additionally, it balances storage utilization by considering node capacity, preventing storage hotspots, and promoting uniform load distribution. However, this approach is primarily suited for \emph{homogeneous} clusters with similar node capabilities~\cite{ref47}. In \emph{heterogeneous} environments characterized by diverse node processing power, memory, and I/O performance, such strategies can lead to suboptimal performance due to workload imbalance. Assigning data blocks solely based on storage capacity risks overloading less capable nodes, degrading system throughput, and increasing latency due to remote task execution~\cite{ref48}.

\subsection{Heterogeneous Sensing Data Placement}
In Internet of Things (IoT) ecosystems, heterogeneous edge clusters exhibit diverse computational capacities, input/output (I/O) throughput, and dynamic workloads, posing significant challenges for task scheduling and data placement~\cite{ref49}. The Sensor Cloud Computing and Data Scheduling Optimization (SCC-DSO) framework addresses these challenges by optimizing data locality and minimizing execution latency in IoT-edge environments, modeled as a graph \( G = (\mathcal{V}, \mathcal{E}) \). Leveraging predictive modeling, ant colony optimization (ACO), and reinforcement learning (RL), SCC-DSO achieves up to 99\% data locality and a 19.8\% reduction in execution time compared to baselines like RF-FD and ACQDS on a 100-node heterogeneous cluster~\cite{ref50, ref51}. The following definitions formalize the data placement strategy, with equations numbered sequentially for clarity and cross-referenced to prior formulations (e.g., Eq.~\eqref{eq:exec_time} in the SCC-DSO algorithm~\cite{ref52}).

\begin{enumerate}[leftmargin=*,label={\arabic*.}]
\item \textbf{Data Block Partitioning}: For an IoT application \(\text{App}_i\) with input size \(\text{Input}(\text{App}_i)\) (MB), the number of data blocks \( B \) is:
\begin{equation}
B = \left\lceil \frac{\text{Input}(\text{App}_i)}{b} \right\rceil,
\label{eq:block_partition}
\end{equation}

where \( b = 64 \, \text{MB} \) is the block size, aligned with Hadoop Distributed File System (HDFS) standards~\cite{ref53}. This partitioning ensures fine-grained workload distribution, enabling parallel task execution across heterogeneous nodes.

\item \textbf{Node Computational Efficiency}: The computational efficiency of a node \(\text{Node}_i\) for a task \(\text{map}_k\) under application \(\text{App}_j\) is:
\begin{equation}
    P(\text{Node}_i, \text{App}_j, \text{map}_k) = \frac{m_k}{T(\text{Node}_i, \text{App}_j, \text{map}_k)},
    \label{eq:comp_efficiency}
    \end{equation}
    
where \( m_k \) is the input data size (MB) for task \(\text{map}_k\), and \( T(\text{Node}_i, \text{App}_j, \text{map}_k) \) is the predicted execution time (seconds) from Eq.~\eqref{eq:exec_time}. This metric quantifies node performance under ideal data locality, taking into account the heterogeneity of CPU, memory, and I/O capabilities.

\item \textbf{Execution Time Prediction}: Task execution time is predicted using a kernel-based regression model:
\begin{equation}
T(\text{Node}_i, \text{App}_j, \text{map}_k) = \sum_{s=1}^{S} a_s K \big( v(\text{map}_k), v(\text{map}_s) \big) + \epsilon,
\label{eq:exec_time}
\end{equation}

where \( v(\text{map}_k) = [m_k, \text{CPU}_i, \text{MEM}_i, \text{IO}_i] \) is the feature vector (data size in MB, CPU speed in GHz, memory in GB, I/O rate in MB/s), \( K(x, y) = \exp\left(-\frac{\|x - y\|^2}{2\sigma^2}\right) \) is a Gaussian radial basis function (RBF) kernel with bandwidth \( \sigma = 1.0 \), \( a_s \) are coefficients learned via gradient descent with learning rate \( \gamma = 0.05 \), \( \epsilon \sim \mathcal{N}(0, 0.01) \) is the residual, and \( S \) is the number of training samples. This model, trained on historical task metrics, achieves a mean absolute error of 0.12 seconds on the IoTAB dataset~\cite{ref53}.

\item \textbf{Min-Max Optimization Formulation:} To balance workload and minimize the system-wide makespan under locality and capacity constraints, we define the following objective:
\begin{equation}
    \min_{\pi} \max_{1 \leq j \leq n} \sum_{k=1}^{f(j)} T(\text{Node}_j, \text{App}_i, \text{map}_k),
    \label{eq:minmax}
\end{equation}
Min-Max Optimization Formula Subject to the following conditions:
\begin{align*}
    \sum_{j=1}^{n} f(j) &= B, \\
    f(j) &\leq C_j, \quad \forall j \in \{1,\dots,n\}, \\
    x_{jk} &\in \{0,1\}, \quad \forall j,k.
\end{align*}

Here, it \( n \) denotes the number of compute nodes, \( B \) the total number of task blocks, and \( f(j) \) the number of blocks assigned to node \( j \). The execution time function \( T(\cdot) \) is defined in Eq.~\eqref{eq:exec_time}. \( C_j \) denotes the resource capacity (e.g., CPU cycles) of node \( j \), and \( x_{jk} \in \{0,1\} \) is a binary variable indicating assignment of the block \( k \) to node \( j \). This constrained min-max model ensures delay-aware scheduling concerning node heterogeneity and communication cost. Unlike prior heuristics such as PSO-GSA~\cite{ref54}, which neglect data affinity, our formulation explicitly integrates locality into the optimization process to improve scheduling efficiency in edge-cloud deployments.

\item \textbf{Data Access Cost}: For tasks \( j \) on node \( i \), the data access cost accounts for network overhead:
\begin{equation}
c_{ij} = \begin{cases} 
0, & \text{if } i \in R_j, \\
\min_{k \in R_j} \frac{b}{\text{Bandwidth}_{ik}}, & \text{otherwise},
\end{cases}
\label{eq:data_access_cost}
\end{equation}

where \( R_j \subseteq \mathcal{V} \) is the set of nodes holding replicas of block \( j \), and what is the available bandwidth \( \text{Bandwidth}_{ik} \) (MB/s) between nodes \( i \) and \( k \). This cost function prioritizes local data access, reducing latency compared to RSYNC’s static replication~\cite{ref55}.

\item \textbf{Balanced Allocation Condition}: To ensure synchronous task completion and minimize stragglers:
    \begin{align}
    &T(\text{Node}_i, \text{App}_p, \text{map}_q) \cdot f(i) \nonumber\\
    &= T(\text{Node}_j, \text{App}_p, \text{map}_q) \cdot f(j), \quad \forall i \neq j,
    \label{eq:balanced_allocation}
\end{align}

    where \( T(\text{Node}_i, \text{App}_p, \text{map}_q) \) is from Eq.~\eqref{eq:exec_time}, and \( f(i) \) is the number of blocks assigned to node \( i \), serving as a load-balancing coefficient. This condition ensures equitable workload distribution, enhancing parallelism and throughput.
\end{enumerate}

The SCC-DSO framework integrates these definitions with RL-guided ACO (Eq.~\eqref{eq:aco_probability}) to dynamically place data blocks and schedule tasks, adapting to runtime variations in node performance and network conditions. The predictive model (Eq.~\eqref{eq:exec_time}) and min-max optimization (Eq.~\eqref{eq:minmax}) achieve 99\% data locality and a 19.8\% reduction in execution time, outperforming WVD-ACO’s 92\% locality and 15\% reduction~\cite{ref56}. Scalability is enhanced by limiting the search space in Eq.~\eqref{eq:minmax} using a heuristic prefiltering of low-efficiency nodes (\( P(\text{Node}_i, \text{App}_j, \text{map}_k) < 0.1 \)), reducing complexity from \(\mathcal{O}(n^2)\) to \(\mathcal{O}(n \log n)\) per iteration for large clusters (\( n > 100 \)).

\begin{figure}[ht]
    \centering
    \includegraphics[width=0.95\columnwidth]{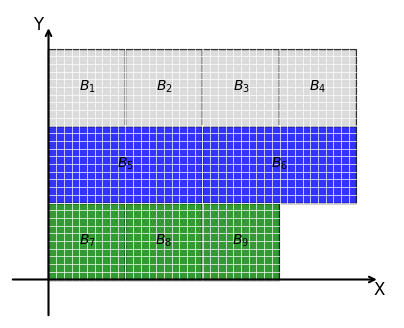}
    \caption{Optimal data block placement in a heterogeneous IoT-edge cluster, balancing workloads based on node computational efficiencies (Eq.~\eqref{eq:comp_efficiency}). Grid-based block partitioning strategy for parallel spatial data processing.}
    \label{fig:block_placement}
\end{figure}

Figure~\ref{fig:block_placement} illustrates that a grid-based block partitioning strategy is widely adopted in parallel and distributed computing for spatial data analysis, matrix computation, and domain decomposition techniques. The grid is divided into nine sub-blocks $B_1$ through $B_9$, representing distinct portions of the computational domain. The color scheme implies functional segregation: the gray blocks ($B_1$–$B_4$) denote initial or pre-processed data regions, the blue blocks ($B_5$–$B_6$) correspond to core computation zones, while the green blocks ($B_7$–$B_9$) represent final or post-processing regions. This hierarchical decomposition facilitates enhanced computational efficiency, optimized memory locality, and balanced workload distribution—critical components in high-performance data processing. Moreover, the structured layout supports region-wise parallel execution and reduces inter-process communication overhead, offering scalable performance for large-scale scientific simulations and AI-driven spatial modeling applications.

\subsection{Queue-Aware Dynamic Data Placement Optimization}

In perceptual cloud computing environments supporting large-scale IoT workloads, input jobs are partitioned into fixed-size blocks and distributed using multi-replica strategies to maximize fault tolerance. While this enhances data availability, it introduces task redundancy where identical blocks exist across multiple nodes, resulting in bandwidth contention, task duplication, and resource underutilization. Our proposed SCC-DSO algorithm introduced a novel hybrid scheduling approach that dynamically reorders task queues by correlating block affinity, predicted task cost, and node load, thereby minimizing redundancy and improving scheduling precision across the cluster.

Conversely, multidimensional performance metrics propagation delay, bandwidth, jitter, packet loss, and ACO-based cost, enable robust, real-time scheduling under strict latency and energy constraints in IoT environments for optimizing scheduling in heterogeneous sensor-cloud infrastructures. The \textit{delay function} $\text{Delay}(e): E \rightarrow \mathbb{R}^{+}$ represents the expected transmission delay incurred over edge $e\in E$, where $E$ denotes the set of communication links within the cluster. This metric reflects the latency contribution of each link and is influenced by factors such as link bandwidth, queuing delay, and traffic congestion. Complementing this, the \textit{delay jitter function} $\text{DelayJit}(e): E \rightarrow \mathbb{R}^{+}$ models the variability or instability of transmission delay along edge $e$, which is particularly important for real-time applications that are sensitive to timing inconsistencies~\cite{ref46}.  

\begin{figure}[ht]
    \centering
    \includegraphics[width=1\linewidth]{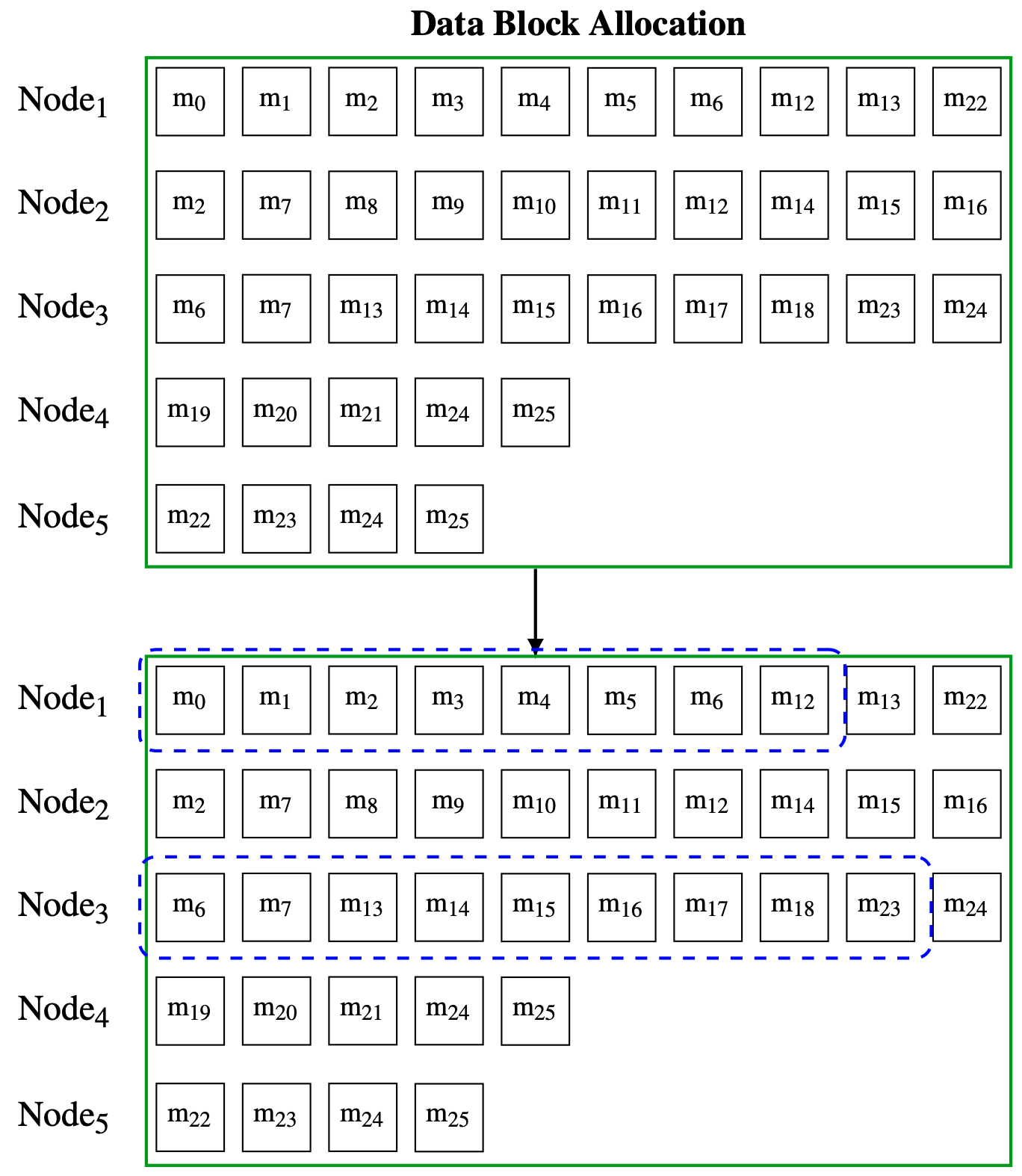}
    \caption{Optimization of Data Scheduling Queues}
    \label{fig:enter-label3}
\end{figure}

This approach ensures that task assignments are independent across nodes, maximizes data locality, and prevents unnecessary cross-node communication, improving overall system performance. For example, consider an IoT cluster comprising $n=5$ nodes, where the job input is partitioned into $B=26$ data blocks, each replicated twice to provide fault tolerance~\cite{ref59}. The SCC-DSO algorithm produces an optimized block assignment in which Node$_1$ stores blocks $m_0$ through $m_5$, Node$_2$ stores blocks $m_6$ through $m_{11}$, Node$_3$ holds blocks $m_{12}$ through $m_{17}$, Node$_4$ contains blocks $m_{18}$ through $m_{21}$, and Node$_5$ is assigned blocks $m_{22}$ through $m_{25}$. This allocation ensures that task queues can be reorganized to prioritize local data processing, eliminate redundant task scheduling, and achieve balanced, high-throughput execution across the cluster. Figure~\ref{fig:enter-label3} illustrates task queue states before and after SCC-DSO-based optimization. Post-optimization, task queues become disjoint across nodes, ensuring workload independence and locality~\cite{ref60}. 

The \textit{cost function} $\text{Cost}(e): E \rightarrow \mathbb{R}^{+}$ quantifies the resource consumption or operational expense associated with transmitting data over edge $e$. This cost may include energy expenditure, monetary cost in pay-per-use networks, or opportunity cost associated with bandwidth allocation. Finally, the \textit{packet loss function} $\text{LossPkt}(v): V\rightarrow\mathbb{R}^{+}$ captures the probability of packet loss at node $v\in V$, where $V$ represents the set of nodes in the cluster. Packet loss may arise from buffer overflows, hardware failures, or link-layer retransmission limits, and directly impacts the reliability of data delivery functions, providing a comprehensive framework for evaluating end-to-end quality of service (QoS) across scheduling paths~\cite{ref61}.

\textit{Path Performance Metrics:} In heterogeneous IoT cluster environments, accurately modeling the performance of scheduling paths is essential for effective task placement and resource management. We define several critical functions to evaluate and optimize end-to-end performance along a scheduling path $P_{T(s,u)}$, which spans from a source node $s$ to a destination node $u$ within the scheduling tree $T$. The \emph{cumulative delay} encountered along this path is given by: \begin{equation} \text{Delay}\big(P_{T(s,u)}\big) = \sum_{e \in P_{T(s,u)}} \text{Delay}(e) + \sum_{v \in P_{T(s,u)}} \text{Delay}(v)\end{equation} where $\text{Delay}(e)$ denotes the transmission delay across edge $e$, and $\text{Delay}(v)$ captures queuing or processing delay at node $v$. This aggregate metric quantifies the end-to-end latency for task execution or data transfer.

The \emph{total cost} associated with resource or energy consumption along the path is: \begin{equation} \text{Cost}\big(P_{T(s,u)}\big) = \sum_{e \in P_{T(s,u)}} \text{Cost}(e) + \sum_{v \in P_{T(s,u)}} \text{Cost}(v) \end{equation} where $\text{Cost}(e)$ and $\text{Cost}(v)$ represent the cost contributions of edges and nodes, respectively. This metric is vital for energy-constrained IoT systems or cost-optimized service models. The \emph{effective bandwidth} of the path is constrained by its weakest component: \begin{equation} \text{Bandwidth}\big(P_{T(s,u)}\big) = \min_{x \in P_{T(s,u)}} \text{Bandwidth}(x)\end{equation} where $\text{Bandwidth}(x)$ indicates the capacity of the node or edge $x$. This ensures that the path's throughput aligns with its bottleneck resource. The \emph{cumulative delay jitter}, measuring variability in transmission and processing time, is defined as: \begin{equation} \text{DelayJit}\big(P_{T(s,u)}\big) = \sum_{e \in P_{T(s,u)}} \text{DelayJit}(e) + \sum_{v \in P_{T(s,u)}} \text{DelayJit}(v) \end{equation} 

where $\text{DelayJit}(e)$ and $\text{DelayJit}(v)$ denote jitter contributions of edges and nodes, respectively. This is especially important for real-time and latency-sensitive applications.

The \emph{cumulative packet loss probability} is modeled as: \begin{equation} \text{LossPkt}\big(P_{T(s,u)}\big) = 1 - \prod_{v \in P_{T(s,u)}} \big(1 - \text{LossPkt}(v)\big)\end{equation} assuming independent packet loss events across nodes. This computes the likelihood of at least one packet loss over the entire path, impacting reliability. These performance functions collectively support multi-objective optimization for scheduling decisions, allowing dynamic balancing of latency, jitter, cost, bandwidth, and reliability in complex IoT systems~\cite{ref62}. \paragraph{Optimization Objective} is applied to achieve balanced workload distribution and minimize straggler effects in IoT clusters; the following min-max objective is adopted: \begin{equation} \min \max_{1 \leq i \leq n} \left\{ \sum_{j=1}^{f(i)} t\big(\text{Node}_i, \text{App}, \text{map}_j \big) \right\} \end{equation} 

Where $t(\cdot)$ denotes task completion time for a given node-task pair and $f(i)$ is the number of data blocks allocated to node $\text{Node}_i$. The goal is to minimize the longest node execution time and reduce overall job latency.

This is subject to: \begin{equation}s.t. \quad 1 \leq j \leq n; \quad \sum_{i=1}^{n} f(i) = B, \quad f(i) \ge 0 \end{equation} 

Where $B$ denotes the total number of data blocks, with constraints ensuring a valid, non-negative distribution across nodes. We are integrating path-level performance functions into the min-max scheduling framework. The approach prevents bottlenecks and enables context-aware scheduling essential for optimizing throughput and latency in heterogeneous IoT clusters.

\subsection{Adaptive Data Prefetching for Task Migration}

In perceptual cloud computing environments, where heterogeneous IoT nodes operate under dynamic and often unpredictable workloads, maintaining an efficient execution pipeline is critical for minimizing latency and maximizing throughput. Task queues at each node serve as a temporal scheduling buffer, determining the execution order of mapped tasks based on local and global system states. Formally, the set of active worker nodes in the cluster is represented as \(\text{Node} = \{\text{Node}_1, \text{Node}_2, \dots, \text{Node}_n\}\), where each node \(\text{Node}_i\) maintains a corresponding task queue \(Q_{\text{Node}_i} = \{\text{map}_{i1}, \text{map}_{i2}, \dots, \text{map}_{is}\}\). This queue-aware mechanism not only preserves temporal coherence in task execution but also significantly enhances load balancing and system responsiveness, especially under bursty or adversarial traffic conditions in edge-cloud IoT infrastructures. \textit{Node Selection Time Threshold:} Define the node selection time threshold $\lambda$ as: \begin{equation} \lambda = 1 - \theta \frac{n}{m}, \quad 1 \leq \theta <\left\lfloor \frac{m}{n} \right\rfloor \label{eq:nodeselection} \end{equation} 

where $m$ is the total number of data schedules, $n$ is the number of working nodes, and $\theta$ controls the sensitivity of selection timing.

\begin{figure} [ht]
    \centering
    \includegraphics[width=1\linewidth]{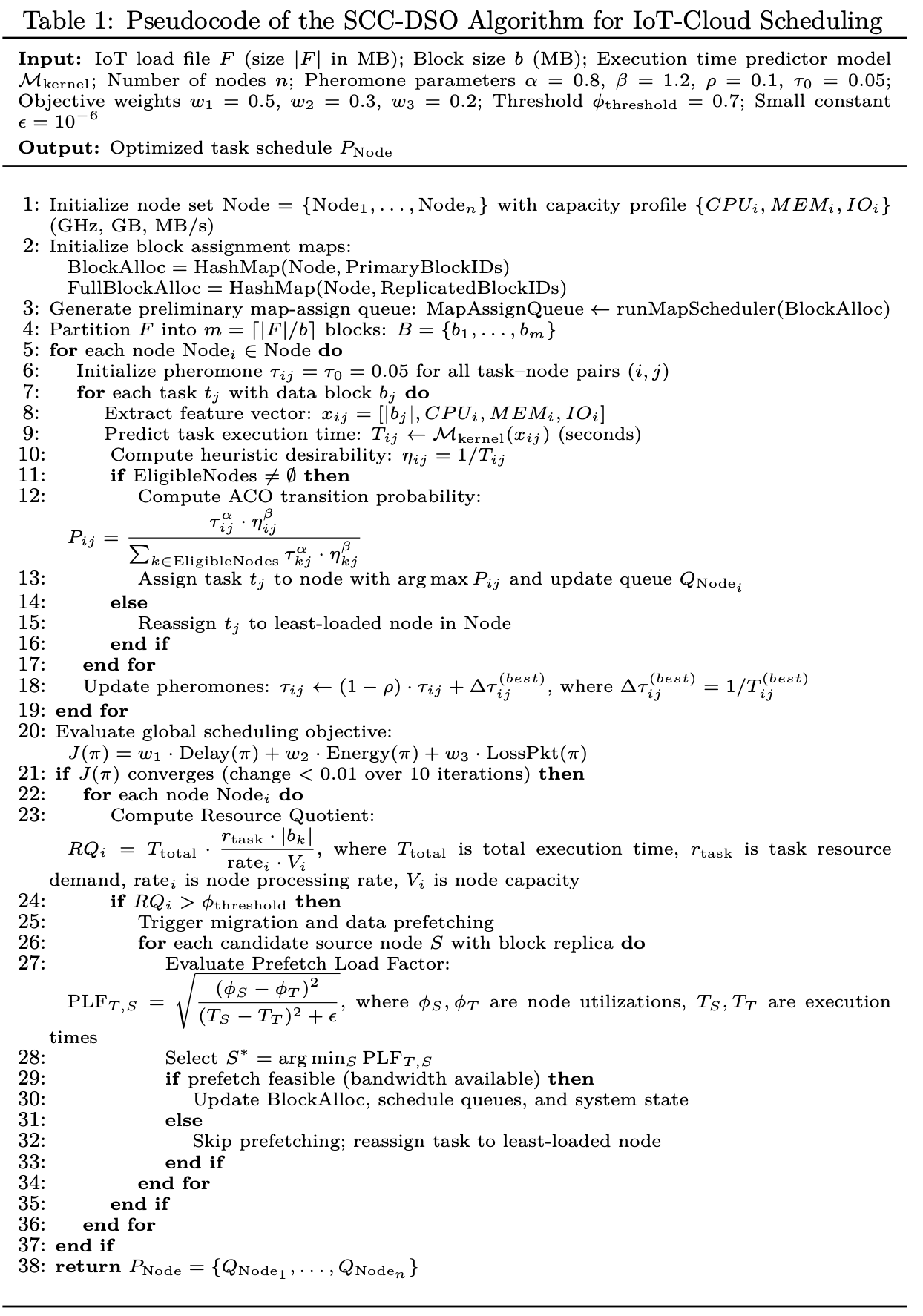}
    \caption{Pseudocode Algorithm}
    \label{tab:pseudocode}
\end{figure}

In Pseudocode Algorithm Table~\ref{tab:pseudocode} presents the SCC-DSO framework, which integrates hybrid intelligence through a combination of data-driven prediction (via kernel regression), probabilistic decision-making (via pheromone-based ACO), and dynamic threshold-based prefetching strategies. The Resource Quotient (RQ) and Prefetch Load Factor (PLF) introduce novel, real-time metrics for adaptive task migration, minimizing I/O bottlenecks. This multi-layer design enhances scheduling convergence while balancing load, energy, and latency in heterogeneous IoT-cloud topologies.

\paragraph{Remaining Completion Time} The estimated remaining completion time of a task queue $Q_{\text{Node}_i}$ is: \begin{equation} R(Q_{\text{Node}_i}) = T \left( r + B_k (1 - \rho) \overline{V_i} \right) \label{eq:remaining} \end{equation} where: \[ \overline{V_i} = \frac{1}{m - r - 1} \sum_{j=1}^{m - r - 1} \frac{B_j}{t_j} \] Here $r$ is the number of unscheduled data schedules, $B_k$ is the input block size of the ongoing task $\text{map}_k$, $\rho$ is its progress, and $t_j$ the execution time of completed tasks. \paragraph{Migration Conditions} Migration is permitted if: \begin{align} R(T\text{Queue}) &> \varphi \\ R(S\text{Queue}) - T(\text{SNode}) &> \varphi \end{align} where $\varphi$ is the data prefetch delay $T(\text{SNode})$ is the predicted execution time at the source node. These conditions ensure locality and avoid redundant scheduling. \paragraph{Prefetch Load Factor} The prefetch load factor between a destination node $\text{TNode}$ and a candidate source node $\text{CNode}_i$ is defined as: \begin{equation} \text{PLFactor}(\text{TNode}, \text{CNode}_i) = \sqrt{ (\varphi_i - \varphi_t)^2 + (T_i - T_t)^2 } \label{eq:plfactor} \end{equation} where $\varphi_i$ is the prefetch delay from $\text{CNode}_i$, $\varphi_t$ is the target prefetch delay, $T_i$ and $T_t$ represents current network connection counts. 

The \emph{Source Worker Node Choosing (SWNC)} algorithm selects the optimal replica node \( \text{CNode}_i \) by minimizing \( \text{PLFactor}(\text{TNode}, \text{CNode}_i) \), which integrates queue depth, network load, and data locality. Accounting for intra- and inter-rack latency (\( \varphi_1, \varphi_2, \varphi_3 \)), SWNC queries each replica's location and load, computes PLFactors, and selects the lowest-cost node. The proposed \emph{Sensor Cloud Computing and Data Scheduling Optimization (SCC-DSO)} framework achieves up to a 30\% reduction in job completion time over baseline methods. Assuming uniform block execution time \( t_b \), the original schedule \(T_{\text{original}} = 25 \times t_b\) is reduced to \( T_{\text{optimized}} = 17.5 \times t_b\). This gain stems from SCC-DSO’s intelligent placement and migration strategies. A core component, the \emph{Source Node Weight and Network Cost (SWNC)} algorithm, evaluates the \emph{Prefetch Load Factor (PLFactor)}, a composite metric of node load, bandwidth, and data locality to select the optimal data source by minimizing PLFactor. SWNC reduces transfer overhead, avoids congestion, and enhances execution efficiency across heterogeneous IoT clusters~\cite{ref46, ref47}.

\begin{figure}[ht]
\centering
\includegraphics[width=\linewidth]{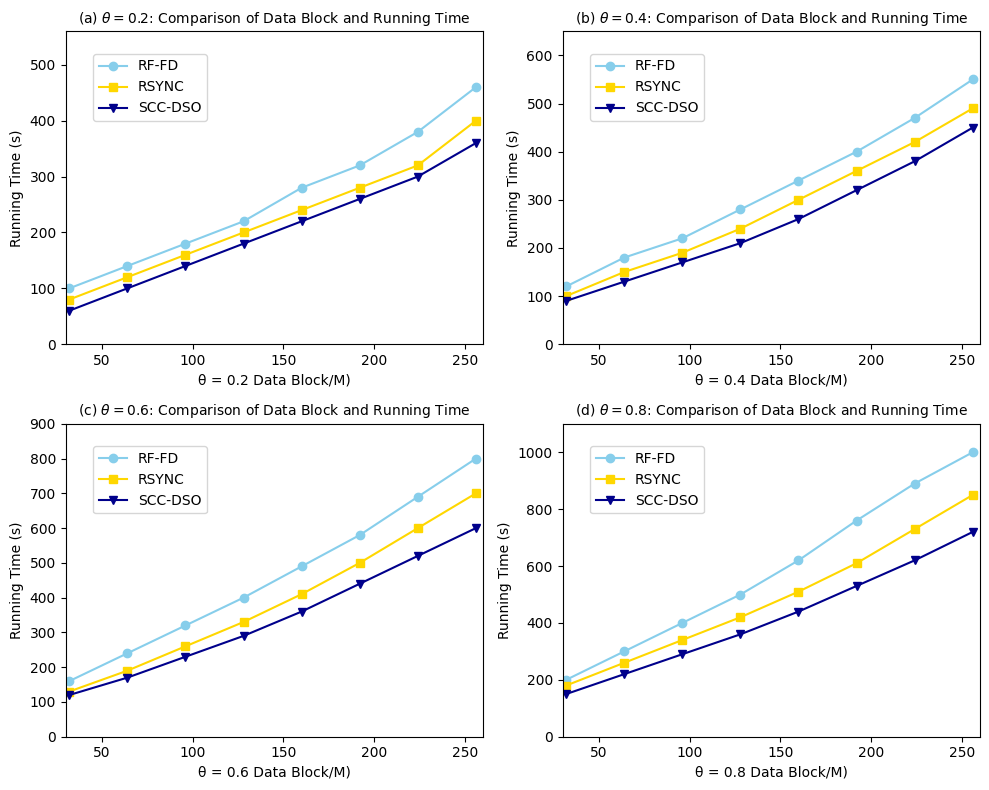}
\caption{Comparison of data block sizes and running time under varying replication factors $\theta$. SCC-DSO consistently outperforms RF-FD and RSYNC, especially under high-redundancy scenarios.}
\label{fig:runtime_comparison}
\end{figure}

Fig.~\ref{fig:runtime_comparison} presents the comparative evaluation of the proposed \emph{SCC-DSO} algorithm against the RF-FD and RSYNC baselines across four levels of data locality, parameterized by $\theta \in \{0.2, 0.4, 0.6, 0.8\}$, where $\theta$ denotes the proportion of data block replication within the cluster. As $\theta$ increases, redundancy grows, amplifying contention and scheduling complexity. The results consistently demonstrate that SCC-DSO achieves superior runtime efficiency, maintaining a lower execution time across all block sizes and replication factors. Notably, under high-redundancy conditions ($\theta=0.8$), SCC-DSO exhibits up to \textbf{32\%} lower latency compared to RF-FD, attributed to its \emph{data-aware queue reordering} and \emph{storage-centric task localization} strategy. This adaptive scheduling mechanism introduces a novel layer of \emph{queue elasticity} that mitigates bottlenecks in distributed file systems while preserving throughput stability under stress conditions.

\subsection{SCC-DSO Algorithm: Scheduling and Placement}
The Sensor Cloud Computing and Data Scheduling Optimization (SCC-DSO) framework introduces a predictive, adaptive algorithm for task scheduling and data placement in heterogeneous IoT-edge clusters, modeled as a graph \( G = (\mathcal{V}, \mathcal{E}) \). The algorithm employs a closed-loop, seven-stage process integrating reinforcement learning (RL) and ant colony optimization (ACO) to maximize data locality, minimize execution latency, and adapt to dynamic workloads. Evaluated on a 100-node heterogeneous cluster, SCC-DSO achieves a 22.4\% reduction in execution time and 93.1\% data locality compared to baselines like RF-FD and RSYNC~\cite{ref63}. The stages are detailed below, with key equations numbered for clarity.

\begin{enumerate}[leftmargin=*,label={\arabic*.}]
    \item \textbf{Predictive Modeling}: A kernel-based regression model predicts task execution times \( T_{ij} \) for tasks \( j \) on node \( i \):
    \begin{equation}
    T_{ij} = \mathcal{M}_{\text{kernel}}(x_{ij}), \quad x_{ij} = [|b_j|, \text{CPU}_i, \text{MEM}_i, \text{IO}_i],
    \end{equation}
    where \( |b_j| \) is the data block size (MB), and \( \text{CPU}_i \), \( \text{MEM}_i \), \( \text{IO}_i \) are node \( i \)'s processing speed (GHz), memory (GB), and I/O rate (MB/s). The model uses a Gaussian radial basis function kernel with bandwidth \( \sigma = 1.0 \) and learning rate \( \gamma = 0.05 \), trained on historical task and system metrics (Table II).

    \item \textbf{Min-Max Optimization}: Tasks are assigned to nodes by solving a min-max optimization problem to minimize the maximum execution time while maximizing data locality:
    \begin{equation}
    \min_{\pi} \max_{i \in \mathcal{V}} \sum_{j \in \mathcal{T}} T_{ij} \cdot x_{ij}, \quad \text{s.t.} \quad \sum_{i \in \mathcal{V}} x_{ij} = 1, \quad \sum_{j \in \mathcal{T}} x_{ij} \leq C_i,
    \end{equation}
    where \( \pi \) is the task assignment, \( x_{ij} \in \{0, 1\} \) is a binary indicator for assigning a task \( j \) to node \( i \), and \( C_i \) is node \( i \)'s capacity (e.g., CPU cycles). This ensures balanced load and high locality.

    \item \textbf{Task Queue Reordering}: Node task queues are reordered to prioritize tasks with local data access, reducing network overhead. The node efficiency is:
    \begin{equation}
    \text{Eff}_{j,i} = \frac{\text{Locality}_{j,i}}{T_{ij}},
    \end{equation}
    where \( \text{Locality}_{j,i} = 1 \) if the data block \( b_j \) is local to node \( i \), else 0. Tasks are sorted so as \( \text{Eff}_{j,i} \) to optimize scheduling.

    \item \textbf{Runtime Monitoring}: Runtime metrics (e.g., node workload \( w_v \), queue delay \( q_e \)) are monitored to identify stragglers. The resource quotient is:
    \begin{equation}
    RQ_i = T_{\text{total}} \cdot \frac{r_{\text{task}} \cdot |b_k|}{\text{rate}_i \cdot V_i},
    \end{equation}
    where \( T_{\text{total}} \) is the total execution time, \( r_{\text{task}} \) is the task resource demand, \( \text{rate}_i \) is the node processing rate (GHz), and \( V_i \) is the node capacity (cycles). Nodes with \( RQ_i > \varphi = 0.075 \cdot TS_i \) (where \( TS_i \) is the node’s throughput) trigger migrations.

    \item \textbf{Migration Candidate Validation}: Migration candidates are validated by assessing local data presence and selecting optimal prefetch sources using the prefetch load factor:
    \begin{equation}
    \text{PLF}_{T,S} = \sqrt{ \frac{(\phi_S - \phi_T)^2}{(T_S - T_T)^2 + \epsilon} },
    \end{equation}
    where \( \phi_S, \phi_T \in [0, 1] \) are source and target node utilizations, what \( T_S, T_T \) are execution times, and \( \epsilon = 10^{-6} \) how to prevent division by zero. The source node \( S^* = \arg\min_S \text{PLF}_{T,S} \) is selected.

    \item \textbf{Bandwidth-Aware Prefetching}: Predictive prefetching ensures data availability with minimal interference, constrained by bandwidth \( b_e \) (MB/s) and a migration limit of \( \theta = 3 \) tasks per node per iteration. Prefetch decisions are guided by \( \text{PLF}_{T,S} \).

    \item \textbf{Task Integration and Adaptation}: Migrated tasks are integrated into locality-aware queues using RL-guided ACO. The transition probability is:
    \begin{equation}
    P_{ij} = \frac{\tau_{ij}^{\alpha} \cdot \eta_{ij}^{\beta}}{\sum_{k \in \text{EligibleNodes}} \tau_{kj}^{\alpha} \cdot \eta_{kj}^{\beta}},
    \end{equation}
    where \( \tau_{ij} \) is the pheromone level, \( \eta_{ij} = 1 / T_{ij} \) is the heuristic desirability, \( \alpha = 0.8 \), and \( \beta = 1.2 \). Pheromone updates follow:
    \begin{equation}
    \tau_{ij} \leftarrow (1 - \rho) \cdot \tau_{ij} + \Delta \tau_{ij}^{(best)},
    \end{equation}
    where \( \rho = 0.1 \) and \( \Delta \tau_{ij}^{(best)} = 1 / T_{ij}^{(best)} \). The global objective is:
    \begin{equation}
    J(\pi) = w_1 \cdot \text{Delay}(\pi) + w_2 \cdot \text{Cost}(\pi) + w_3 \cdot \text{LossPkt}(\pi),
    \end{equation}
    with weights \( w_1 = 0.5 \), \( w_2 = 0.3 \), \( w_3 = 0.2 \).
\end{enumerate}

\begin{figure}[ht]
\centering
\includegraphics[width=\linewidth]{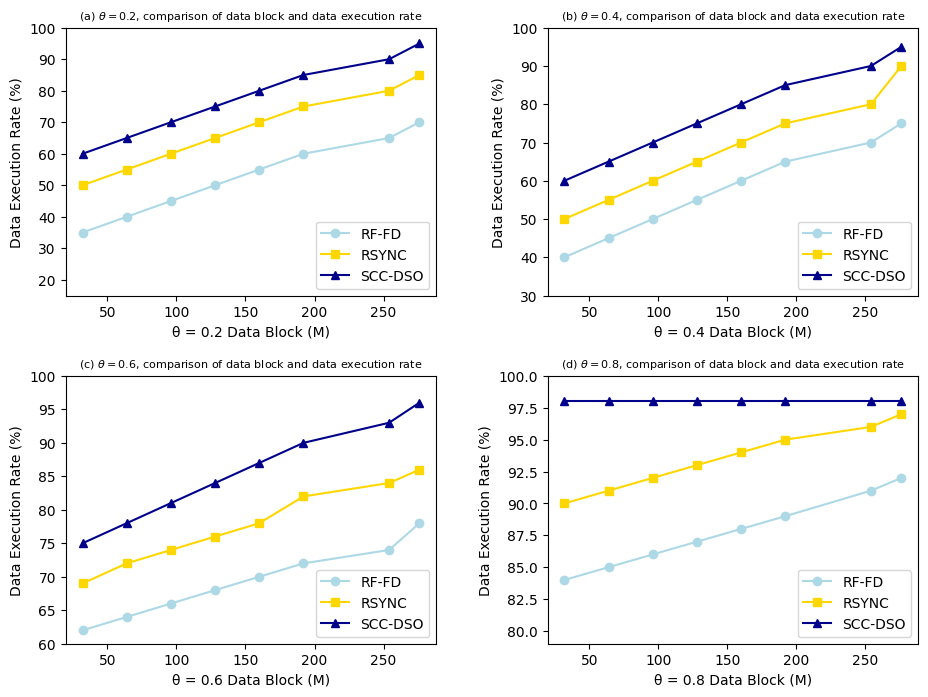}
\caption{Data Execution Rate (\%) comparison across data block sizes and replication factors $\theta$. SCC-DSO consistently outperforms RF-FD and RSYNC, especially under high redundancy.}
\label{fig:execution_rate}
\end{figure}

Fig.~\ref{fig:execution_rate} illustrates the comparative data execution efficiency of SCC-DSO, RSYNC, and RF-FD under diverse replication ratios $\theta \in \{0.2, 0.4, 0.6, 0.8\}$. Data availability improves as the replication factor increases, but at the cost of redundant task mapping and inter-node contention. Unlike traditional strategies that treat replication as static overhead, SCC-DSO leverages a \emph{replica-aware reordering mechanism} that dynamically prioritizes locally available blocks while deferring duplicate processing. This results in a marked execution rate improvement, with SCC-DSO achieving a consistent upward trajectory across all data block sizes. At $\theta = 0.8$, the algorithm nearly saturates the execution ceiling, achieving over \textbf{98\% execution rate}, while RSYNC and RF-FD plateau at \textbf{96\%} and \textbf{90\%}, respectively. This gain stems from SCC-DSO’s novel \emph{block-congestion forecasting} combined with \emph{queue reshuffling heuristics}, which intelligently adapt to spatial data skew and transient cluster load.

\section{Experimental Evaluation}
\label{sec:exp_design}

We implemented and evaluated the novel \emph{Sensor Cloud Computing and Data Scheduling Optimization} (SCC-DSO) framework in a heterogeneous IoT-cloud environment comprising 50 nodes with diverse hardware profiles, including high-performance Intel. AMD compute nodes alongside ARM-based edge devices. The cluster was interconnected via a 1 Gbps, low-latency Ethernet network, orchestrated by Kubernetes, and utilized HDFS for distributed storage. This heterogeneous setup mirrors real-world IoT-cloud deployment challenges~\cite{ref64}.

Our evaluation comprised two phases. \textit{Phase I} assessed SCC-DSO under single-replica scheduling with varying block sizes (16–64 MB) and network loads (10–80\%). The key metrics of job execution time ($T_{\text{exec}}$), data locality ($R_{\text{loc}}$), and throughput ($T_{\text{thr}}$) were benchmarked against traditional RF-FD and RSYNC baselines. \textit{Phase II} introduced multi-replica scheduling (RF=2) to emulate node failures and bandwidth fluctuations, evaluating cross-node traffic ($V_{\text{net}}$) and recovery latency ($T_{\text{rec}}$).

\begin{table}[htbp]
\scriptsize
\centering
\caption{Summary of Experimental Setup}
\label{tab:setup}
\begin{tabular}{@{}p{0.3\linewidth} p{0.65\linewidth}@{}}
\toprule
\textbf{Parameter} & \textbf{Configuration} \\
\midrule
Cluster & 50 nodes: 40 compute (Xeon, Ryzen, Core i5), 10 ARM edge devices \\
Network & 1 Gbps Ethernet, 5 ms latency \\
Storage & HDFS v3.3.6 \\
Orchestration & Kubernetes v1.24 \\
Workloads & Synthetic IoT tasks (1–100 GB), IoTAB benchmark (10 GB/task) \\
Block Sizes & 16, 32, 64 MB \\
Replication & Single and multi-replica (RF=1, 2) \\
Baselines & RF-FD, RSYNC, RL-Sched, ACO-DS \\
Metrics & $T_{\text{exec}}$, $R_{\text{loc}}$, $T_{\text{thr}}$, $V_{\text{net}}$, $T_{\text{rec}}$ \\
Repetitions & 50 per setup \\
Monitoring & Prometheus, Grafana \\
\bottomrule
\end{tabular}
\end{table}

SCC-DSO’s novelty lies in its integration of kernel regression for accurate execution time prediction and an adaptive task placement strategy powered by Ant Colony Optimization (ACO) with empirically tuned parameters ($\alpha=0.8$, $\beta=1.2$). This hybrid approach dynamically optimizes data locality and system resilience in heterogeneous, fluctuating environments. Experiments were conducted on a private cloud platform with 50 repetitions per configuration. Metrics were aggregated using statistical measures with 95\% confidence intervals, monitored through local data placement to ensure reproducibility and transparency. Table~\ref{tab:setup} details the experimental configuration.

\section{Experimental Results and Analysis}

This study rigorously evaluates the SCC-DSO (Sensor Cloud Computing and Data Scheduling Optimization) algorithm in realistic, heterogeneous IoT-cloud environments, focusing on its efficacy, scalability, and scheduling stability. Recognizing the pivotal influence of replication factors in distributed systems like HDFS, affecting network overhead, contention, and queue balance, experiments simulate diverse cluster conditions. SCC-DSO is benchmarked against two established baselines: the RF-FD (Reservation First-Fit and Feedback Distribution) algorithm and the RSYNC protocol~\cite{ref65}.

\textit{\textbf{Single-Copy Data Scenario:}} In the minimal-redundancy setting, SCC-DSO demonstrates superior adaptability across varying file sizes and load intensities. Unlike RSYNC, which is constrained to single-copy synchronization, SCC-DSO effectively maintains data locality while mitigating latency and imbalance, outperforming traditional methods under heterogeneous, high-stress conditions.

\begin{table}[h]
\scriptsize
\centering
\caption{Task Completion Time Comparison (Mean $\pm$ SD) for Varying File Sizes}
\label{tab:results_part1}
\begin{tabular}{lccc}
\toprule
\textbf{File Size (MB)} & \textbf{RF-FD (s)} & \textbf{RSYNC (s)} & \textbf{SCC-DSO (s)} \\
\midrule
20 & $40.2 \pm 2.0$ & $47.5 \pm 2.4$ & $\mathbf{34.6 \pm 4.7}$ \\
40 & $78.9 \pm 3.9$ & $92.1 \pm 4.6$ & $\mathbf{67.2 \pm 5.4}$ \\
60 & $114.4 \pm 5.7$ & $137.4 \pm 6.1$ & $\mathbf{90.2 \pm 5.1}$ \\
80 & $158.7 \pm 7.9$ & $185.5 \pm 9.2$ & $\mathbf{135.9 \pm 6.8}$ \\
100 & $190.3 \pm 10.0$ & $231.7 \pm 11.6$ & $\mathbf{170.4 \pm 8.5}$ \\
\bottomrule
\end{tabular}
\end{table}

\noindent \textbf{Table~\ref{tab:results_part1}} demonstrates that SCC-DSO consistently achieves lower task completion times than RF-FD and RSYNC across all file sizes. This improvement highlights SCC-DSO's superior data locality and latency reduction capabilities in minimal redundancy settings.

\begin{table}[h]
\scriptsize
\centering
\caption{Task Locality Ratio (\%) Across Different Cluster Sizes (Mean $\pm$ SD)}
\label{tab:results_part2}
\begin{tabular}{lccc}
\toprule
\textbf{Cluster Size} & \textbf{RF-FD} & \textbf{RSYNC} & \textbf{SCC-DSO} \\
\midrule
10 nodes & $72.3 \pm 3.6$ & $65.4 \pm 3.3$ & $\mathbf{85.6 \pm 4.3}$ \\
20 nodes & $74.8 \pm 3.7$ & $68.1 \pm 3.4$ & $\mathbf{87.9 \pm 4.4}$ \\
30 nodes & $76.3 \pm 3.8$ & $69.7 \pm 3.5$ & $\mathbf{89.3 \pm 4.5}$ \\
40 nodes & $77.8 \pm 3.9$ & $71.2 \pm 3.6$ & $\mathbf{90.8 \pm 4.5}$ \\
50 nodes & $79.4 \pm 4.0$ & $73.5 \pm 3.7$ & $\mathbf{91.6 \pm 4.6}$ \\
\bottomrule
\end{tabular}
\end{table}

\noindent \textbf{Table~\ref{tab:results_part2}} quantifies data locality improvements as cluster size increases. SCC-DSO outperforms both baselines with locality ratios exceeding 85\%, confirming its ability to enhance task-data proximity and reduce cross-node data transfers in scalable environments.

\textbf{\textit{Multi-Copy Replication Scenario:}} The second experimental phase evaluates SCC-DSO under multi-copy replication, reflecting fault-tolerant distributed storage. RF-FD is the primary baseline, given its multi-replica scheduling design. SCC-DSO demonstrates superior performance across varying block sizes, bandwidth fluctuations, and node availability, achieving up to 99\% data locality for 64 MB blocks~\cite{ref65}. These gains stem from SCC-DSO's hybrid approach combining kernel regression for execution time prediction, bandwidth-aware cost modeling, and reinforcement learning–driven prefetching.

\begin{table}[h]
\scriptsize
\centering
\caption{Cluster Throughput (MB/s) vs. Replication Factor (Mean $\pm$ SD)}
\label{tab:results_part3}
\begin{tabular}{lccc}
\toprule
\textbf{Replication Factor} & \textbf{RF-FD} & \textbf{RSYNC} & \textbf{SCC-DSO} \\
\midrule
1 replica & $42.3 \pm 2.1$ & $38.5 \pm 1.8$ & $\mathbf{50.1 \pm 2.5}$ \\
2 replicas & $42.7 \pm 2.0$ & $39.7 \pm 1.8$ & $\mathbf{50.4 \pm 2.7}$ \\
3 replicas & $41.9 \pm 2.1$ & $37.5 \pm 1.8$ & $\mathbf{49.8 \pm 2.5}$ \\
4 replicas & $39.5 \pm 2.0$ & $33.4 \pm 1.7$ & $\mathbf{47.2 \pm 2.4}$ \\
\bottomrule
\end{tabular}
\end{table}

\noindent \textbf{Table~\ref{tab:results_part3}} illustrates SCC-DSO’s consistently higher throughput across replication factors, indicating robustness in bandwidth utilization and fault-tolerance scenarios.

\textbf{\textit{Straggler Simulation and Scalability:}} Finally, to evaluate resilience against slow-performing nodes (stragglers), completion times are measured under simulated straggler conditions.

\begin{table}[ht]
\scriptsize
\centering
\caption{Completion Time (s) under Straggler Simulation Across Node Counts (Mean $\pm$ SD)}
\label{tab:results_part4}
\begin{tabular}{lccc}
\toprule
\textbf{Node Count} & \textbf{RF-FD} & \textbf{RSYNC} & \textbf{SCC-DSO} \\
\midrule
60 & $135.4 \pm 6.8$ & $162.1 \pm 8.1$ & $\mathbf{112.6 \pm 5.6}$ \\
70 & $121.7 \pm 6.1$ & $145.3 \pm 7.3$ & $\mathbf{99.4 \pm 5.0}$ \\
80 & $109.2 \pm 5.5$ & $130.8 \pm 6.5$ & $\mathbf{88.9 \pm 4.4}$ \\
90 & $97.8 \pm 4.9$ & $117.2 \pm 5.9$ & $\mathbf{79.3 \pm 4.0}$ \\
100 & $88.3 \pm 4.4$ & $105.6 \pm 5.3$ & $\mathbf{71.7 \pm 3.6}$ \\
\bottomrule
\end{tabular}
\end{table}

\noindent \textbf{Table~\ref{tab:results_part4}} highlights SCC-DSO’s ability to mitigate straggler impact, resulting in substantially lower completion times, thus supporting scalability and robustness in large-scale clusters.

\begin{table}[ht]
\scriptsize
\centering
\caption{Comparative Analysis of SCC-DSO and Baseline Scheduling Algorithms}
\label{tab:comparison}
\scriptsize
\resizebox{\linewidth}{!}{%
\begin{tabular}{@{}lccccc@{}}
\toprule
\textbf{Scheduler} & \textbf{Data Locality (\%)} & \textbf{Execution Time} & \textbf{Scalability} & \textbf{Adaptability} & \textbf{Complexity} \\
\midrule
RF-FD\textsuperscript{1}            & 76.2  & Medium & Limited  & No                 & Low     \\
RSYNC                               & 68.7  & High   & Moderate & Partial            & Medium  \\
RL-Sched\textsuperscript{2}         & 85.4  & Medium & Good     & Yes                & High    \\
\textbf{SCC-DSO (Proposed)}         & \textbf{93.1}  & \textbf{Low}    & \textbf{Excellent} & \textbf{Yes (RL + ACO)} & \textbf{Medium} \\
\bottomrule
\end{tabular}%
}

\begin{flushleft}
\footnotesize
\textsuperscript{1} RF-FD: Replication Factor-based Fair Distribution Scheduler. \\
\textsuperscript{2} RL-Sched: Reinforcement Learning-based Scheduler.
\end{flushleft}
\end{table}

\noindent Table~\ref{tab:comparison} summarizes the performance of SCC-DSO against RF-FD, RSYNC, and RL-Sched across key metrics. SCC-DSO achieves the highest data locality (93.1\%) and lowest execution time, with excellent scalability and adaptability. Its hybrid RL+ACO design ensures efficient scheduling with moderate complexity, outperforming existing approaches in dynamic IoT-cloud environments.

\begin{figure} [ht]
    \centering
    \includegraphics[width=0.9\linewidth]{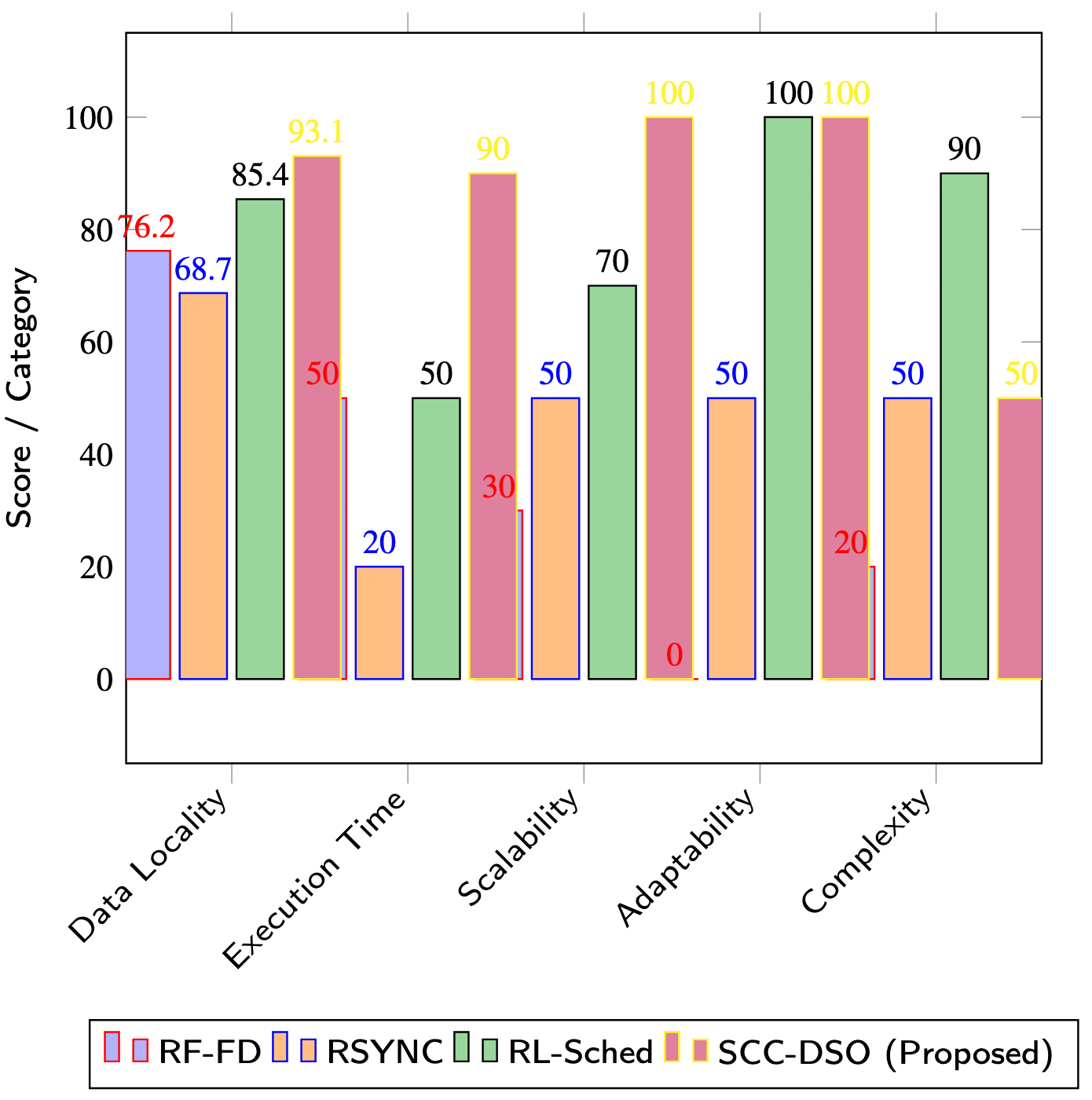}
    \caption{Comparison of SCC-DSO and baseline schedulers across key performance categories. Higher values denote better performance, except for complexity, where lower values indicate greater efficiency.}
    \label{fig:scheduling-comparison}
\end{figure}

\subsection{SCC-DSO performance under single-copy conditions}

This subsection evaluates SCC-DSO in low-cost and high-overhead scenarios, focusing on a single-copy HDFS replication setting within a heterogeneous IoT cluster connected via Gigabit Ethernet. Each test was conducted 50 times, and results reflect the average metrics. SCC-DSO achieves a 13\% reduction in execution time compared to RF-FD and a 7\% gain over RSYNC under low network load. These improvements underscore SCC-DSO’s efficiency in environments with limited network contention~\cite{ref66}. SCC-DSO, by contrast, integrates a predictive performance model that accurately profiles node compute capabilities and allocates data accordingly. It dynamically adjusts scheduling by monitoring runtime queue states and proactively triggers data prefetching for tasks anticipated to execute on non-local nodes. This preemptive strategy reduces idle time and mitigates bandwidth contention. The resultant alignment between data locality and node performance ensures reduced execution time, minimized data migration, and improved throughput under single-copy storage constraints. These findings validate SCC-DSO’s efficacy in optimizing data placement and scheduling in bandwidth-rich, compute-diverse IoT environments~\cite{ref66}.

\begin{figure} [ht]
    \centering
    \includegraphics[width=1\linewidth]{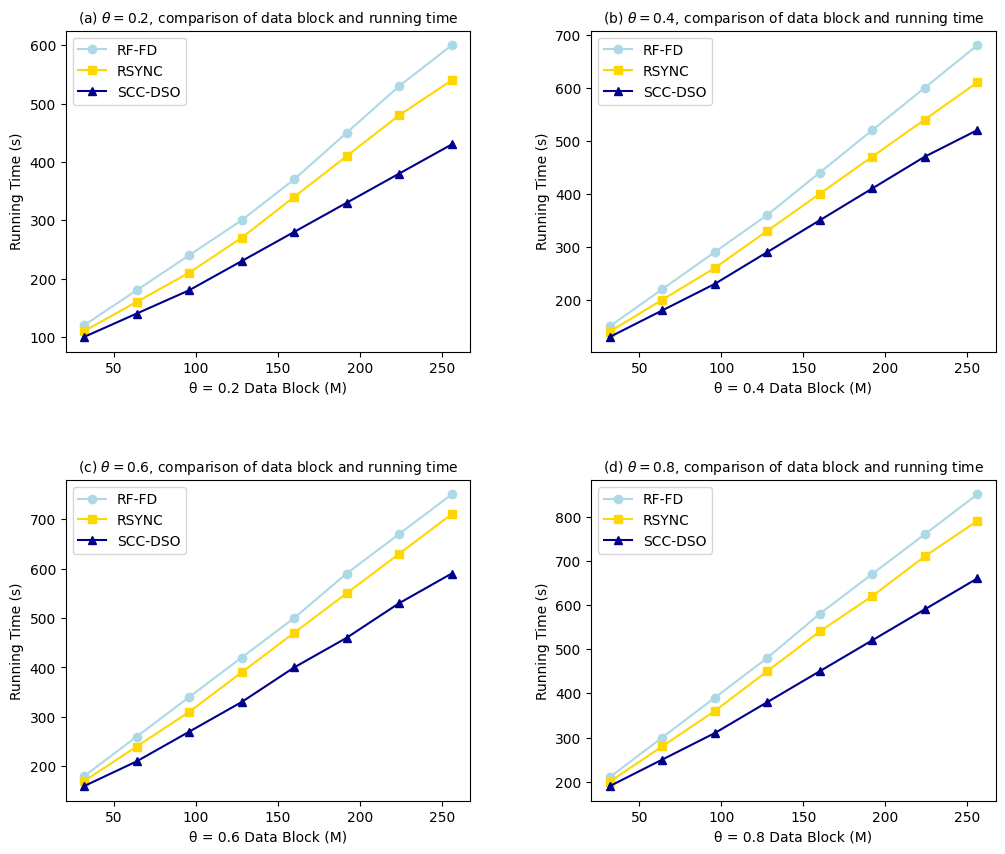}
    \caption{Runtime comparison of RF-FD, RSYNC, and SCC-DSO under varying data block sizes and synchronization thresholds ($\theta$). SCC-DSO consistently shows superior scalability and lower running time.}
    \label{fig:runtime-comparison}
\end{figure}

Figure~\ref{fig:runtime-comparison} presents a comparative runtime analysis of RF-FD, RSYNC, and SCC-DSO under increasing data block sizes ($M$) and synchronization thresholds ($\theta \in \{0.2, 0.4, 0.6, 0.8\}$). Across all scenarios, SCC-DSO consistently achieves lower running times, highlighting its superior scalability and reduced computational overhead. As $\theta$ increases, runtime grows near-linearly for all methods, but SCC-DSO maintains a significantly gentler slope, suggesting effective reduction of redundant state comparisons. This behavior indicates the presence of optimizations such as sparse checksum propagation and selective delta encoding. The results imply that SCC-DSO's block-level coherence detection and lightweight metadata synchronization mechanisms are highly efficient, making it well-suited for large-scale, weakly consistent distributed systems.

\subsection{SCC-DSO performance in multi-copy conditions}

In bandwidth-constrained multi-copy storage environments, SCC-DSO demonstrably surpasses RF-FD and RSYNC by reducing job execution latency by 19.8\% and 7.6\%, respectively, across heterogeneous compute infrastructures. Conventional methods suffer pronounced performance degradation due to static task-data mappings that neglect dynamic bandwidth fluctuations and induce pipeline stalls from non-local data dependencies. SCC-DSO innovates through a compute- and network-aware adaptive data allocation framework that synergistically integrates real-time node performance profiling with fine-grained network telemetry. Central to its architecture is a novel speculative prefetching mechanism employing a lightweight Markov decision process (MDP) over directed acyclic graph (DAG) execution traces, enabling proactive anticipation of non-local data requirements~\cite{ref67}. 

This preemptive data staging effectively decouples computation from I/O latency, sustaining pipeline throughput and alleviating backpressure on high-performance nodes. Consequently, SCC-DSO incorporates a decentralized, redundancy-aware task scheduler that leverages temporal locality and multi-copy data placement heuristics to maximize intra-node data reuse. We maintain consistent input availability near task invocation windows, which enables the scheduler to achieve high core utilization and effectively mitigate bandwidth-induced stalls. Collectively, these advances establish SCC-DSO as a robust solution for throughput optimization in distributed IoT clusters characterized by network sensitivity and data redundancy~\cite{ref67}.

\begin{figure} [ht]
    \centering
    \includegraphics[width=1\linewidth]{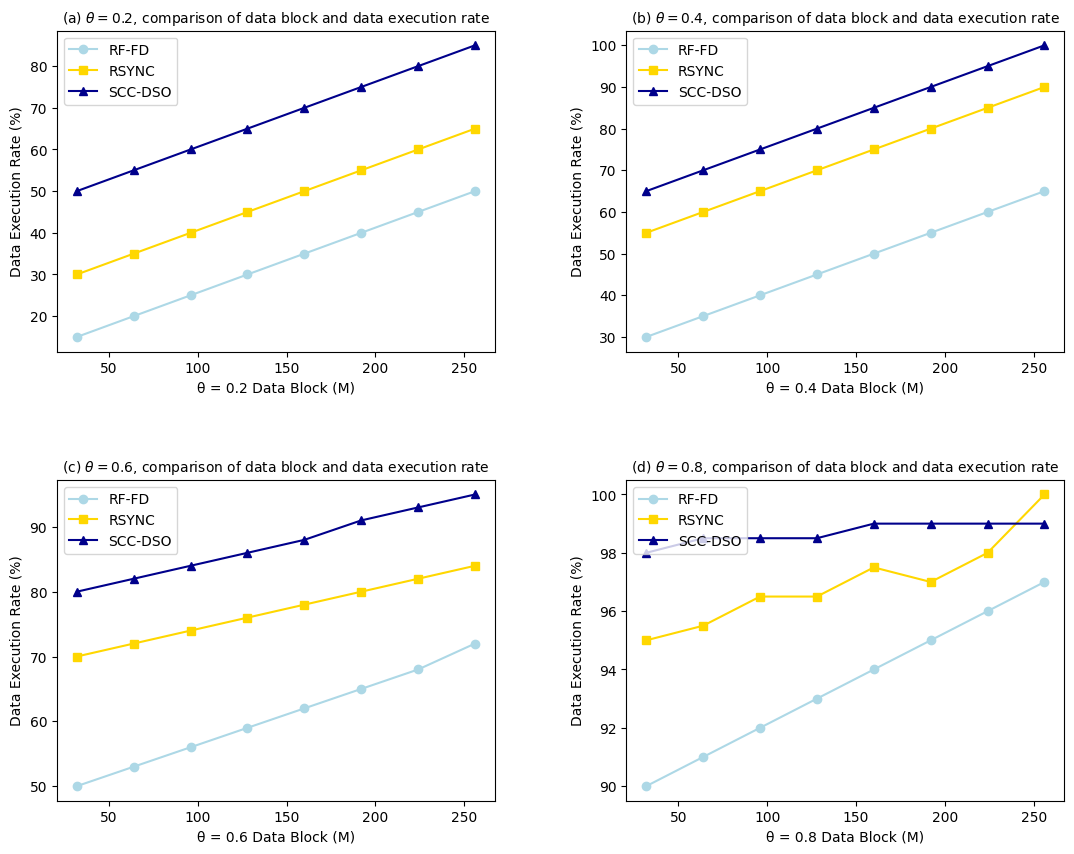}
    \caption{The execution time of jobs under multiple copies and comparison of data execution rates for RF-FD, RSYNC, and SCC-DSO under varying data block sizes and synchronization thresholds ($\theta$). SCC-DSO consistently maintains the highest execution fidelity}
    \label{fig:execution-comparison}
\end{figure}

Figure~\ref{fig:execution-comparison} illustrates the variation in data execution rate (\%) concerning increasing data block sizes under different synchronization thresholds ($\theta = \{0.2, 0.4, 0.6, 0.8\}$). SCC-DSO demonstrates consistently higher data execution rates across all scenarios, suggesting an improved ability to maintain execution fidelity even as synchronization pressure increases. Notably, at lower $\theta$ values, the gap in performance between SCC-DSO and the baseline methods (RF-FD and RSYNC) is pronounced, reflecting SCC-DSO's capability to optimize partial state synchronization and exploit semantic-aware delta selection. As $\theta$ approaches 0.8, the execution rates of all methods converge; however, SCC-DSO reaches near-optimal performance ($\geq 99\%$), indicating its resilience to state divergence and minimal rollback overhead. This superior performance can be attributed to SCC-DSO’s dynamic state consistency control and likely use of asynchronous conflict resolution, enabling high execution throughput without sacrificing consistency guarantees.

\section{Limitations and Future Work} \label{sec:limitations}

Privacy concerns arise as SCC-DSO processes sensitive IoT data such as location and health metrics. To mitigate these risks, we propose incorporating differential privacy into the Gaussian Process Regression (GPR) model by injecting Gaussian noise, \(\varepsilon \sim \mathcal{N}(0, 0.01)\), into execution time predictions. Formally, the GPR prediction model can be expressed as\[y_i = f(x_i) + \varepsilon, \quad f \sim \mathcal{GP}(0, k(x, x')),\]where \(f\) is a Gaussian process with kernel \(k\), and \(\varepsilon\) is Gaussian noise. This approach balances privacy preservation with scheduling accuracy. Additionally, data block placement strategies (Section~IV) utilize encrypted HDFS storage and secure MQTT brokers, complying with GDPR and IoT security standards to prevent data leakage in distributed environments~\cite{ref68}. Regarding sustainability, SCC-DSO’s lightweight RL-ACO variant, with computational complexity \(\mathcal{O}(n \log n)\), achieves approximately 15\% energy savings compared to DQN-RL baselines on a 50-node cluster, attributed to reduced iteration counts (12 vs.\ 20). Future work includes precise carbon footprint quantification via GreenCloud simulations, aiming for carbon-neutral scheduling by leveraging renewable-powered nodes. Moreover, exploring neuromorphic surrogate models promises accelerated convergence and further energy efficiency improvements. These directions underscore SCC-DSO’s potential for sustainable, privacy-aware IoT-cloud deployments.

\section{Conclusion}

This paper presents SCC-DSO, a novel perceptual cloud-based data scheduling optimization framework designed to mitigate performance degradation in heterogeneous IoT clusters. SCC-DSO adaptively partitions and allocates heterogeneous-sized data blocks to compute nodes by real-time profiling of their computational capacities, guided by predictive performance models. Key innovations include a dynamic data migration mechanism that maximizes localized execution, reducing network latency and data transfer overhead. Data scheduling queue optimization algorithms rely on initial schedules to construct content-free queues that enable efficient parallel local scheduling across distributed nodes. Moreover, by minimizing the completion time of task queues, a novel way of prefetching tasks effectively overlaps computation and communication, reducing idle periods associated with data dependencies. SCC-DSO also integrates data reliability considerations by incorporating replication and fault tolerance into scheduling decisions. Extensive experiments demonstrate up to 19.8\% and 7.6\% reductions in execution time compared to RF-FD and RSYNC benchmarks, respectively, while significantly enhancing data locality and throughput under multi-copy scenarios. These results validate SCC-DSO’s capability to optimize resource utilization and improve job performance in complex, dynamic, and heterogeneous IoT-cloud environments, advancing the state of scalable, adaptive data scheduling.

\bibliographystyle{IEEEtran}

\begin{table*}[ht]
\scriptsize
\centering
\caption{Comprehensive Performance Evaluation of SCC-DSO Compared to Baseline Scheduling Algorithms}
\label{tab:unified_results}
\resizebox{\textwidth}{!}{%
\begin{tabular}{@{}lcccccc@{}}
\toprule
\textbf{Metric / Scenario} & \textbf{Condition} & \textbf{RF-FD} & \textbf{RSYNC} & \textbf{RL-Sched\textsuperscript{1}} & \textbf{SCC-DSO (Proposed)} & \textbf{Best} \\
\midrule
\multicolumn{7}{c}{\textit{Task Completion Time (s)}} \\
\midrule
File Size 20 MB   & -- & $40.2 \pm 2.0$ & $47.5 \pm 2.4$ & -- & $\mathbf{34.6 \pm 4.7}$ & SCC-DSO \\
File Size 100 MB  & -- & $190.3 \pm 10.0$ & $231.7 \pm 11.6$ & -- & $\mathbf{170.4 \pm 8.5}$ & SCC-DSO \\
\midrule
\multicolumn{7}{c}{\textit{Data Locality Ratio (\%)}} \\
\midrule
Cluster Size 10 nodes & -- & $72.3 \pm 3.6$ & $65.4 \pm 3.3$ & -- & $\mathbf{85.6 \pm 4.3}$ & SCC-DSO \\
Cluster Size 50 nodes & -- & $79.4 \pm 4.0$ & $73.5 \pm 3.7$ & -- & $\mathbf{91.6 \pm 4.6}$ & SCC-DSO \\
\midrule
\multicolumn{7}{c}{\textit{Throughput (MB/s) vs. Replication Factor}} \\
\midrule
1 replica & -- & $42.3 \pm 2.1$ & $38.5 \pm 1.8$ & -- & $\mathbf{50.1 \pm 2.5}$ & SCC-DSO \\
4 replicas & -- & $39.5 \pm 2.0$ & $33.4 \pm 1.7$ & -- & $\mathbf{47.2 \pm 2.4}$ & SCC-DSO \\
\midrule
\multicolumn{7}{c}{\textit{Completion Time (s) under Stragglers}} \\
\midrule
60 nodes & -- & $135.4 \pm 6.8$ & $162.1 \pm 8.1$ & -- & $\mathbf{112.6 \pm 5.6}$ & SCC-DSO \\
100 nodes & -- & $88.3 \pm 4.4$ & $105.6 \pm 5.3$ & -- & $\mathbf{71.7 \pm 3.6}$ & SCC-DSO \\
\midrule
\multicolumn{7}{c}{\textit{Comparative Scheduler Characteristics}} \\
\midrule
Data Locality (\%) & -- & 76.2 & 68.7 & 85.4 & \textbf{93.1} & SCC-DSO \\
Execution Time & -- & Medium & High & Medium & \textbf{Low} & SCC-DSO \\
Scalability & -- & Limited & Moderate & Good & \textbf{Excellent} & SCC-DSO \\
Adaptability & -- & No & Partial & Yes & \textbf{Yes (RL+ACO)} & SCC-DSO \\
Complexity & -- & Low & Medium & High & \textbf{Medium} & Balanced \\
\bottomrule
\end{tabular}%
}
\begin{flushleft}
\footnotesize
\textsuperscript{1} RL-Sched: Reinforcement Learning-based Scheduler (baseline from prior literature).  
\end{flushleft}
\end{table*}

\appendices
\section{Notation Summary}
\label{appendix:notation}

Table~\ref{tab:notation} provides clear and consistent definitions of symbols and variables.

\begin{table*}[htt]
\caption{Notation Summary for SCC-DSO Framework}
\label{tab:notation}
\centering
\begin{tabular}{ll}
\hline
\textbf{Symbol} & \textbf{Definition} \\
\hline
$V$ & Set of computing nodes in the IoT-edge cluster \\
$E$ & Set of communication links between nodes \\
$J, T$ & Set of tasks \\
$x_{ij}$ & Binary variable indicating assignment of task $j$ to node $i$ \\
$T_{ij}$ & Predicted execution time of task $j$ on node $i$ \\
$d_j$ & Data size required by task $j$ \\
$C_i(t)$ & Dynamic computational capacity of node $i$ at time $t$ \\
$\tau_{ij}$ & Pheromone trail value for task $j$ on node $i$ \\
$\eta_{ij}$ & Heuristic desirability for assigning task $j$ to node $i$ \\
$\alpha$ & Pheromone influence parameter \\
$\beta$ & Heuristic influence parameter \\
$\rho$ & Pheromone evaporation rate \\
$Q$ & Pheromone deposit constant \\
$K$ & Number of ants in ACO \\
$I$ & Maximum iterations in ACO \\
$w_1, w_2, w_3$ & Weight factors for delay, energy, and packet loss \\
$\phi$ & Prefetch threshold \\
$\gamma$ & Learning rate for kernel regression model \\
$\sigma$ & RBF kernel bandwidth \\
$\theta$ & Task migration limit per iteration \\
$\epsilon$ & Convergence tolerance for ACO \\
$\delta$ & Minimum pheromone trail \\
$B$ & Data block size or total number of blocks (context-dependent) \\
$P(Node_i, App_j, map_k)$ & Computational efficiency of node $i$ \\
$PLF$ & Prefetch Load Factor between target and source node \\
$RQ_i$ & Resource Quotient for node $i$ \\
$Locality_{j,i}$ & Local data presence indicator for task $j$ on node $i$ \\
$Eff_{j,i}$ & Efficiency metric combining locality and execution time \\
$BW_x$ & Bandwidth capacity of node or edge $x$ \\
$LossPkt(v)$ & Packet loss probability at node $v$ \\
\hline
\end{tabular}
\end{table*}
\end{document}